\documentclass[12pt]{article}
\pdfoutput=1
\usepackage{subfigure}
\usepackage{cancel}
\usepackage{cite}
\usepackage{lipsum}
\usepackage{amsmath}
\usepackage{makecell}
\usepackage{indentfirst}
\usepackage{multirow}
\usepackage{amssymb}
\usepackage{feynmp,color}
\usepackage[utf8]{inputenc}
\usepackage{hyperref}

\usepackage{amsmath,amssymb,amsthm,amsxtra,overpic,bbm,bm,epsfig,ulem}
\begin{document}
\unitlength=1mm
\begin{center}
		{\Large \bf The signatures of the new particles $h_{2}$ and $Z^{}_{\mu\tau}$  at e-p colliders in the ${U(1)}_{L^{}_{\mu} - L^{}_{\tau}  }$ model}
\end{center}

\vspace{0.05cm}

\begin{center}
{\bf Jin-Xin Hou \footnote{E-mail: houjinxin\_email@yeah.net},  Chong-Xing Yue \footnote{E-mail: cxyue@lnnu.edu.cn} } \\ %Yu-chen Guo \footnote{E-mail: lgguoyuchen@126.com}} \\
{Department of Physics, Liaoning Normal University, Dalian 116029, China}
\end{center}

\begin{abstract}
	
Considering the superior performances of the future e-p colliders, LHeC and FCC-eh, we discuss the feasibility of detecting the extra neutral scalar $h_{2}$ and the  light gauge boson $Z^{}_{\mu\tau}$, which are predicted by the ${U(1)}_{L^{}_{\mu} - L^{}_{\tau}}$ model. Taking into account the experimental constraints on the relevant free parameters, we consider all possible production channels of $h_{2}$ and $Z^{}_{\mu\tau}$ at e-p colliders and further investigate their observability  through the optimal channels in the case of the beam polarization P($e^{-}$)= -0.8. We find that the signal significance above 5$\sigma$ of $h_{2}$ as well as $Z^{}_{\mu\tau}$ detecting can be achieved via $e^{-}p\to{e^{-}jh_{2}(\to{Z_{\mu\tau}Z_{\mu\tau}})}\to~e^{-}j+\cancel{E}^{}_{T}$ process and a 5$\sigma$ sensitivity of $Z^{}_{\mu\tau}$ detecting can be gained via $e^{-}p\to{e^{-}jh_{1}(\to{Z^{}_{\mu\tau}Z^{}_{\mu\tau}})\to}~e^{-}j+\cancel{E}^{}_{T}$ process at e-p colliders with appropriate parameter values and a designed integrated luminosity. However, the signals of $h_{2}$ decays into pair of SM particles are difficult to be detected.
	
\end{abstract}

\vspace{0.2cm}
\section*{\uppercase\expandafter{\romannumeral1}. Introduction}

As is known, the standard model (SM) of particle physics is one of the most successful theories over the past decades which describes a variety of experimental results over the wide range of energy scale from eV to TeV. Discovery of the 125 GeV Higgs boson at the Large Hadron Collider (LHC) in 2012~\cite{higgs1,Chatrchyan:2012xdj}  proves the success of the SM once again. However, so far  the SM still has certain limitations. Some experimental facts have been plaguing people, and there is an urgent need to extend the SM. For instance, the sub-eV masses and peculiar mixing pattern of neutrinos ~\cite{Fukuda:1998mi,Ahmad:2001an}, the muon $(g-2)$ anomalous magnetic moment~\cite{Ade:2013zuv}, the exploration of dark matter (DM)~\cite{Bertone:2004pz} and dark energy~\cite{Perlmutter:1998np,Riess:1998cb}, the baryon asymmetry of the Universe and so on. Furthermore, discovery of Higgs boson provides an outstanding portal to new physics (NP) beyond the SM. Precision measurements of the Higgs boson properties are also one of the most important tasks of high-energy particle physics due to its possible role as portal to beyond the SM (BSM) sectors~\cite{Cakir:2009xi,Liang:2010gm,Zhang:2015ado,Antusch:2016ejd,Curtin:2017bxr}.

So far, there are many well motivated extensions of the SM, such as SUSY~\cite{Porod:2018gli,Carta:2018qke,Mamuzic:2018utn,Kitazawa:2018zys}, two Higgs doublet model~\cite{Babu:2018uik,Kon:2018vmv,Kang:2018jem,Chaber:2018cbi,Li:2018aov,Azevedo:2018llq}, and extension of the SM with an extra $U(1)$ gauge group~\cite{Asai:2018ocx,Allanach:2018vjg,Chauhan:2018uuy,DelleRose:2018eic,Das:2019pua,Das:2015nwk,Das:2016zue}. In this work, we will consider the gauged ${U(1)}_{L^{}_{\mu}- L^{}_{\tau}}$ extension of the SM due to its relatively simple theoretical structure, which has a complete gauge group $SU(3)_C \times SU(2)_L \times U(1)_Y \times {U(1)}_{L^{}_{\mu} - L^{}_{\tau} }$ and is called the ${U(1)}_{L^{}_{\mu} - L^{}_{\tau}}$ model~\cite{He:1991qd,He:1990pn,Drees:2018hhs,Hou:2019jxv,Banerjee:2018mnw,Nomura:2018yej,Biswas:2016yjr,Biswas:2016yan,Chun:2018ibr}. One of the advantages of the ${U(1)}_{L^{}_{\mu} - L^{}_{\tau}}$ model is that the anomaly cancellation does not require any extra chiral fermionic degrees of freedom. In this model, the breaking of ${L^{}_{\mu}- L^{}_{\tau}}$ symmetry conduces to additional terms in the neutrino mass matrix, which offers an explanation for the neutrino masses and mixing simultaneously~\cite{Biswas:2016yjr,Biswas:2016yan}. Besides, the scalar sector has been expanded by two additional complex scalar singlets ($\varphi_{H}$ and $\varphi_{DM}$) with nonzero ${L^{}_{\mu}- L^{}_{\tau}}$ charge. The scalar $\varphi_{DM}$ can act as a viable DM candidate~\cite{Araki:2017wyg}. The other scalar $\varphi_{H}$ acquires a vacuum expectation value (VEV) $v_{\mu\tau}$ and thereby making an additional neutral scalar $h_{2}$ after spontaneous breaking of ${U(1)}_{L^{}_{\mu} - L^{}_{\tau}}$, which indicates that $h_{2}$ has a mass of the same order with $v^{}_{\mu\tau}$  about 10 GeV - 1000 GeV~\cite{Nomura:2018yej,Gninenko:2018tlp}. In principle, the additional neutral scalar can be produced and decay via their mixing with the SM-like Higgs boson $h_{1}$~\cite{DelleRose:2018ndz}.

On the other hand, an extra neutral gauge boson $Z^{}_{\mu\tau}$ is also introduced and obtains a mass after spontaneous symmetry breaking of ${U(1)}_{L^{}_{\mu} - L^{}_{\tau}}$. $Z^{}_{\mu\tau}$ does not couple to the SM quarks and the first generation leptons, which makes it avoid restrictions coming from lepton and hadron colliders such as LEP and LHC. Therefore, the mass of $Z^{}_{\mu\tau}$ can be as light as 100 MeV for a low value of gauge coupling $g^{}_{\mu\tau} \leq 10^{-3}$, which is required to meet the limits arising from neutrino trident production. The $Z^{}_{\mu\tau}$ with an MeV-scale mass can resolve the muon $(g-2)$ anomaly, explain the deficit of cosmic neutrino flux~\cite{Biswas:2016yjr,Kamada:2015era,Araki:2015mya} and resolve the problem of relic abundance of DM in the scenario with a light weakly interacting massive particle~\cite{Baek:2008nz,Baek:2015fea,Patra:2016shz} simultaneously. Therefore, searching for its possible collider evidences  plays a vital role in exploring NP. Many attempts to discover this kind of new particles have been made in the meson decay experiment~\cite{Banerjee:2017hhz}, beam dump experiment~\cite{Anelli:2015pba}, electron-positron collider experiments~\cite{Abe:2010gxa} and so on.

Searches for the new particles predicted by the ${U(1)}_{L^{}_{\mu} - L^{}_{\tau}}$ model are presently being conducted at the LHC and ILC~\cite{Nomura:2018yej}. While, another Higgs factory besides the LHC and ILC, such as the LHeC (Large Hadron electron Collider) and FCC-eh (Future Circular Collider in hadron-electron mode)~\cite{DelleRose:2018ndz,Bordry:2018gri,Han:2018rkz,Das:2018usr}, could precisely determine their specific properties. In this paper, we mainly devote to study of the $h_{2}$ and $Z^{}_{\mu\tau}$ productions and further explore the possibility of detecting their signatures at e-p colliders. We present a full simulation study of the production cross sections of $h_{2}$ and $Z^{}_{\mu\tau}$ with the beam polarization P($e^{-}$)= -0.8. Then, we investigate their observability through the processes $e^{-}p\to{e^{-}jh_{2}(\to{Z^{}_{\mu\tau}Z^{}_{\mu\tau}})\to}~e^{-}j+\cancel{E}^{}_{T}$, $e^{-}p\to{e^{-}jh_{1}(\to{Z^{}_{\mu\tau}Z^{}_{\mu\tau}})\to}~e^{-}j+\cancel{E}^{}_{T}$ and $e^{-}p\to{\nu j}{h^{}_{ 2}(\to{ZZ})\to}~ 2l^{+}2l^{-}j+\cancel{E}^{}_{T}$, respectively. We further analyze the signal significance of $h_{2}$ and $Z_{\mu\tau}$ detecting which depends on the free parameters. Our numerical results show that the signals of $h_{2}\to Z^{}_{\mu\tau}Z^{}_{\mu\tau}$ and $h_{1}\to Z^{}_{\mu\tau}Z^{}_{\mu\tau}$ are promising to be detected at e-p colliders with appropriate parameter values and high integrated luminosity. But, due to the interference of substantial backgrounds and the low number of events, searching for the signal of the decay channel $h_{2}\to ZZ$ are harder to achieve at e-p colliders.

Rest of the paper has been arranged in the following manner. In Sec.~II, we briefly review the basic features of the ${U(1)}_{L^{}_{\mu}-L^{}_{\tau}}$ model and show the allowed parameter space of this model.  In Sec.~III, we not only give the partial widths of the main decay channels of the scalar $h_{2}$, but also calculate its production cross sections via the $W^{+}W^{-}$ and $ZZ$ fusion processes. The production cross sections of the new gauge boson $Z^{}_{\mu\tau}$ through $h_{2}$ and $h_{1}$ decays are calculated in Sec.~IV. We estimate the numbers of the signal and background events, and investigate the signal observability and discovery potentiality of $h_{2}$ and $Z^{}_{\mu\tau}$ through their respective promising production channels in Sec.~V. Finally, our conclusions are given in Sec.~VI.

\section*{\uppercase\expandafter{\romannumeral2}. The Basic Features of the ${U(1)}_{L^{}_{\mu}-L^{}_{\tau}}$  Model}

The gauged ${U(1)}_{L^{}_{\mu} - L^{}_{\tau}  }$ extension of the SM is one of the most extensively studied NP models, which can successfully solve the origin of tiny neutrino masses, the DM relic abundance and the muon $(g-2)$ anomalous magnetic moment. Refs.\cite{Biswas:2016yjr,Biswas:2016yan} have made a detailed analysis about solving these  puzzles in the ${U(1)}_{L^{}_{\mu} - L^{}_{\tau}  }$ model. In this model, the gauge sector of the SM is enhanced by imposing a local ${U(1)}_{L^{}_{\mu} - L^{}_{\tau}  }$ symmetry to the SM Lagrangian, where $L_{\mu}$ and $L_{\tau}$ are the muon and tau lepton numbers, respectively. Therefore, the complete gauged group is $SU(3)_C \times SU(2)_L \times U(1)_Y \times {U(1)}_{L^{}_{\mu} - L^{}_{\tau} }$. The SM particle content  has been extended by including three extra right-handed (RH) neutrinos and two SM gauge singlet scalars. All particles included in the ${U(1)}_{L^{}_{\mu} - L^{}_{\tau}}$ model and their charge assignments under various symmetry groups are listed in Table~\ref{table1}.

\begin{table*}[!ht]
	\caption{\footnotesize Particle contents and corresponding charge assignments under various symmetry groups in the ${U(1)}_{L^{}_{\mu} - L^{}_{\tau}}$ model.}
	\centering
	\begin{tabular}{c|c|c|c|c|c|c|c|c|c|c|c|c}
		\hline
		\hline
		\multirow{2}{*}{Gauge Group} &\multicolumn{3}{|c|}{ Scalar Fields}&\multicolumn{9}{|c}{  Lepton Fields}\\
		\cline{2-13}                   &${\varphi}^{}_{h}$&${\varphi}^{}_{H}$&${\varphi}^{}_{DM}$&$L_{e}$&$L_{\mu}$& $L_{\tau}$&$e_{R}$&$\mu_{R}$&$\tau_{R}$&$N^{e}_{R}$  & $N^{\mu}_{R}$ & $N^{\tau}_{R}$\\
		\hline $ SU(2)_L $       &2   & 1    &  1  & 2  & 2 &2&1&1&1&1&1&1 \\
		\hline $ U(1)_Y $  & 1/2    & 0  & 0     &-1/2 &-1/2&-1/2&-1&-1&-1&0&0&0 \\
		\hline ${U(1)}_{L^{}_{\mu} - L^{}_{\tau}  }$ &0  &  1   & $n_{\mu\tau}$ &0&1&-1&0&1&-1&0&1&-1\\
		\hline
		\hline
	\end{tabular}
	\label{table1}
\end{table*}
\begin{table*}[!ht]
	\centering
	\begin{tabular}{c|c|c|c|c|c|c|c|c|c|c|c|c}
		\hline
		\hline
		\multirow{2}{*}{Gauge Group}  &\multicolumn{12}{|c}{Baryon Fields}\\ \cline{2-13}
		&$u^{}_{L}$&$d^{}_{L}$&$c^{}_{L}$&$s^{}_{L}$& $t^{}_{L}$&$b^{}_{L}$&$u^{}_{R}$&$d^{}_{R}$&$c^{}_{R}$ &$s^{}_{R}$ & $t^{}_{R}$ &$b^{}_{R}$\\
		\hline $ SU(2)_L $       &2   & 2    &  2 & 2  & 2 &2&1&1&1&1&1&1 \\
		\hline $ U(1)_Y $  & 1/6   &1/6  & 1/6     &1/6&1/6&1/6&2/3&-1/3&2/3&-1/3&2/3&-1/3\\
		\hline ${U(1)}_{L^{}_{\mu} - L^{}_{\tau}  }$ &0  &  0  & 0 &0&0&0&0&0&0&0&0&0 \\
		\hline
		\hline
	\end{tabular}
\end{table*}

The Lagrangian of the ${U(1)}_{L^{}_{\mu} - L^{}_{\tau}  }$ model is as follows
\begin{eqnarray}
\mathcal{L} =\!\!\! &\mathcal{L}&\!\!\!\!\!^{}_{SM}+ \mathcal{L}^{}_{N}+\mathcal{L}^{}_{DM}+\left| {D}^{}_{\nu}{\varphi}^{}_{H} \right|^{2}-V-\frac{1}{4}{F}^{\rho\sigma}_{\mu\tau}{F}^{}_{\mu\tau\rho\sigma}\;.
\label{eq1}
\end{eqnarray}
 In above equation we have ignored  the kinetic-mixing term between the groups  $U(1)_{Y}$ and ${U(1)}_{L^{}_{\mu} - L^{}_{\tau}  }$ in the case of assuming that the mixing is very small. The terms $\mathcal{L}^{}_{SM}$, $\mathcal{L}^{}_{N}$ and $\mathcal{L}^{}_{DM}$ represent the SM, right hand (RH) neutrino and DM sectors, respectively. Since the processes we are studying do not involve RH neutrinos, its specific form is not given here.  $\mathcal{L}^{}_{DM}$ represents the dark sector Lagrangian including the kinetic term of the DM candidate ${\varphi}^{}_{DM}$ and the interaction terms of ${\varphi}^{}_{DM}$ with the scalars fields ${\varphi}^{}_{h}$ and ${\varphi}^{}_{H}$. The expression of $\mathcal{L}^{}_{DM}$ is given by
\begin{eqnarray}
\mathcal{L}^{}_{DM}=({D}^{\nu}_{}{\varphi}^{}_{DM})^{\dagger}({D}^{}_{\nu}{\varphi}^{}_{DM})-{\mu}^{2}_{DM}{\varphi}^{\dagger}_{DM}{\varphi}^{}_{DM}-{\lambda}^{}_{DM}({\varphi}^{\dagger}_{DM}{\varphi}^{}_{DM})^2\nonumber\\
-{\lambda}^{}_{Dh}({\varphi}^{\dagger}_{DM}{\varphi}^{}_{DM})({\varphi}^{\dagger}_{h}{\varphi}^{}_{h})-{\lambda}^{}_{DH}({\varphi}^{\dagger}_{DM}{\varphi}^{}_{DM})({\varphi}^{\dagger}_{H}{\varphi}^{}_{H})\;,\quad\quad\;\;
\end{eqnarray}
where the parameters $\lambda^{}_{DM}$, $\lambda^{}_{Dh}$ and $\lambda^{}_{DH}$ are quartic couplings of the scalar
fields.  As these  couplings are feeble ($\sim10^{-12}$)~\cite{Biswas:2016yjr}, the DM can not attain thermal equilibrium with the thermal soup, which is called the Feebly Interacting Massive Particle (FIMP). In Eq.~(\ref{eq1}), the covariant derivatives involving in the kinetic energy term $\left| {D}^{}_{\nu}{\varphi}^{}_{H} \right|^{2}$ of the extra Higgs singlet ${\varphi}^{}_{H}$ can be expressed in a generic form  $D^{}_{\nu}\phi=(\partial^{}_{\nu}+ig^{}_{\mu\tau}Q^{}_{\mu\tau}(\phi)Z^{}_{\mu\tau\nu})\phi$, where $\phi$ is any SM single field which has ${U(1)}_{L^{}_{\mu}-L^{}_{\tau}}$ charge $Q^{}_{\mu\tau}(\phi)$ (listed in Table~\ref{table1}) and  $g^{}_{\mu\tau}$ represents ${U(1)}_{L^{}_{\mu}-L^{}_{\tau}}$ group's gauge coupling constant. The scalar potential $V$ contains all the self interactions of $\varphi_{H}$ and its interactions with SM Higgs doublet. Its expression form is given by
\begin{eqnarray}
\quad\quad V={\mu}^{2}_{H}{\varphi}^{\dagger}_{H}{\varphi}^{}_{H}+{\lambda}^{}_{H}({\varphi}^{\dagger}_{H}{\varphi}^{}_{H})^2+{\lambda}^{}_{hH}({\varphi}^{\dagger}_{h}{\varphi}^{}_{h})
({\varphi}^{\dagger}_{H}{\varphi}^{}_{H})\;.
\label{eq3}
\end{eqnarray}
 The last term in Eq.~(\ref{eq1}) represents the kinetic term for the additional gauge boson $Z^{}_{\mu\tau}$ in terms with field strength tensor  $F^{\rho\sigma}_{\mu\tau}=\partial^{\rho}_{}Z^{\sigma}_{\mu\tau}-\partial^{\sigma}_{}Z^{\rho}_{\mu\tau} $ of the ${U(1)}_{L^{}_{\mu}-L^{}_{\tau}}$ gauge group. When the scalar field $\varphi_{H}$ has a non-zero of VEV, the  ${U(1)}_{L^{}_{\mu} - L^{}_{\tau}  }$ symmetry breaks spontaneously and consequently the corresponding new gauge boson $Z^{}_{\mu\tau}$ obtains the mass ${M}_{Z^{}_{\mu\tau}}= g^{}_{\mu\tau}v^{}_{\mu\tau}$. The
SM Higgs doublet ${\varphi}^{}_{h}$ and the new scalar ${\varphi}^{}_{H}$ take the following form
\begin{eqnarray}
\!\!\!\!\!\varphi^{}_{h} = \left(\begin{array}{c}
\tilde{H}  \cr
\frac{v+H+iA}{ \sqrt2}  \cr
\end{array}\right)\;,&&\! \! \! \!
\varphi^{}_{H} = \left(\begin{array}{c}
\frac{v^{}_{\mu\tau}+H^{}_{\mu\tau}+ia}{\sqrt2}  \cr
\end{array}\right)\;,
\end{eqnarray}
where $\tilde{H}$, $A$ and $a$ are  the massless Nambu-Goldstone Bosons (NGBs) absorbed by the gauge bosons $W^{\pm}$, $Z$ and $Z^{}_{\mu\tau}$, while $v$ and $v^{}_{\mu\tau}$ are the VEVs of the scalars $\varphi^{}_{h}$ and $\varphi^{}_{H}$, respectively.  Furthermore, $H$ and $H^{}_{\mu\tau}$ represent the physical CP-even scalar bosons. When both $\varphi_{h}$ and $\varphi_{H}$ obtain their respective VEVs, there will be a mass mixing between the states $H$ and $H_{\mu\tau}$. The square of scalar mass matrix with off-diagonal elements proportional to $\lambda_{hH}$ is given by
\begin{eqnarray}
\quad\mathcal{M}^{2}_{scalar} = \left(\begin{array}{cc}
2{\lambda}^{}_{h}v^2 & {\lambda}^{}_{hH}v^{}_{\mu\tau}v  \cr
{\lambda}^{}_{hH}v^{}_{\mu\tau}v  & 2{\lambda}^{}_{H}v^{2}_{\mu\tau}  \cr
\end{array}\right)\;.
\label{eq5}
\end{eqnarray}
Rotating the basis states $H$ and $H_{\mu\tau}$ by a suitable angle $\alpha$, we can make the above mass matrix diagonal. The new basis states ($h_{1}$ and $h_{2}$), now representing two physical states, are the linear combinations of $H$ and $H_{\mu\tau}$ with the mixing angle $\alpha$ between $H$ and $H_{\mu\tau}$, which can be expressed as
\begin{eqnarray}
\quad \quad h^{}_{1} =  H\cos{\alpha}+H^{}_{\mu\tau}\sin{\alpha}\;, \quad   h^{}_{2} =  -H\sin{\alpha}+H^{}_{\mu\tau}\cos{\alpha}\;,
\end{eqnarray}
\begin{eqnarray}
\quad\quad\tan 2 \alpha = \frac{ {\lambda}^{}_{hH}v^{}_{\mu\tau} v }{ {\lambda}^{}_{h}v^{2} - {\lambda}^{}_{H}v^{2}_{\mu\tau} } \;.\quad\quad
\end{eqnarray}
When $\alpha\ll1$, $h_{1}$ can be identified as the SM-like Higgs boson which has already been discovered by the CMS~\cite{higgs1} and ATLAS~\cite{Chatrchyan:2012xdj} collaborations in 2012. $h_{2}$ is a new scalar particle. The masses of these two physical scalars $h_{1}$ and $h_{2}$ are given by
\begin{eqnarray}
M^{2}_{h_1} &=& \sqrt{v^2 v_{\mu\tau}^2 \left(\lambda _{hH}^2-2 \lambda _h \lambda_H\right)+\lambda _h^2 v^4+\lambda _H^2 v_{\mu\tau}^4}+\lambda _h v^2+\lambda_H v_{\mu\tau}^2 \;, \nonumber \\
M^{2}_{h_2} &=& -\sqrt{v^2 v_{\mu\tau}^2 \left(\lambda _{hH}^2-2 \lambda _h \lambda_H\right)+\lambda _h^2 v^4+\lambda _H^2 v_{\mu\tau}^4}+\lambda _h v^2+\lambda_H v_{\mu\tau}^2 \;.
\end{eqnarray}
 In this paper, we will assume that the values of $M_{h_{1}}$ and $v$ are fixed at 125 GeV and 246 GeV, respectively.

On one hand, compared with the SM Higgs boson, the couplings of $h_{1}$ with the SM particles are suppressed by a factor $\cos\alpha$ in the ${U(1)}_{L^{}_{\mu} - L^{}_{\tau}  }$ model. Some relevant couplings of the SM-like Higgs boson $h_{1}$ with the SM particles, the new gauge boson $Z^{}_{\mu\tau}$ and the  DM candidate ${\varphi}^{}_{DM}$ are given by
\begin{eqnarray}
\quad\quad\quad\quad\quad g^{}_{Z^{}_{\mu\tau}Z^{}_{\mu\tau}h_{1}}= \frac{2{M}^{2}_{Z^{}_{\mu\tau}}} {v^{}_{\mu\tau}}\sin\alpha\;,\quad
g^{}_{f\bar{f}h_{1}} =-\frac{{M}^{}_{f}} {v}\cos \alpha\;,\quad
g^{}_{VVh_{1}} =\frac{2{M}^{2}_{V}}{v}\cos\alpha\;,\nonumber\quad\quad\quad\\  g^{}_{\varphi^{\dagger}_{DM}{\varphi}^{}_{DM}h_{1}} =-(v\lambda^{}_{Dh}\cos\alpha+v_{\mu\tau}\lambda^{}_{DH}\sin\alpha)\;,\quad\quad\quad\quad\quad\quad\quad\quad\quad\quad\quad\quad\quad\quad\quad
\label{eq9}
\end{eqnarray}
where $f$ represents all of the SM fermions and $V$ represents the electroweak gauge bosons $W^{\pm}$ or $Z$.
On the other hand, similar with $h_{1}$,  the scalar $h_{2}$ can couple to all the SM particles and  other new particles, such as new gauge boson $Z^{}_{\mu\tau}$ and the  DM  particle ${\varphi}^{}_{DM}$. Here, we also list the $h_{2}$ couplings, which are related our calculation
\begin{eqnarray}
\nonumber
g^{}_{Z^{}_{\mu\tau}Z^{}_{\mu\tau}h_{2}}= \frac{2{M}^{2}_{Z^{}_{\mu\tau}}} { v^{}_{\mu\tau}}\cos\alpha\;, \quad
g^{}_{f\bar{f}h_{2}} =\frac{{M}^{}_{f}} {v}\sin \alpha\;, \quad g^{}_{VVh_{2}} =-\frac{2{M}^{2}_{V}}{v}\sin\alpha\;,\;\\ g^{}_{\varphi^{\dagger}_{DM}{\phi}^{}_{DM}h_{2}} =(v\lambda^{}_{Dh}\sin\alpha-v_{\mu\tau}\lambda^{}_{DH}\cos\alpha)\;,\;\;\;\quad\quad\quad\quad\quad\quad\quad\quad\quad\quad\quad\;\;\;\; \nonumber\\
\quad\quad g^{}_{h_{1}h_{1}h_{2}} = 6v\lambda_{h}\cos^{2}\alpha\sin\alpha-6v_{\mu\tau}\lambda_{H}\sin^{2}\alpha\cos\alpha-2v\lambda_{hH}\sin\alpha\quad\quad\quad\quad\quad\nonumber\\+6v\lambda_{hH}\sin^{3}\alpha-v_{\mu\tau}\lambda_{hH}\cos\alpha+3v_{\mu\tau}\lambda_{hH}\sin^{2}\alpha\cos\alpha\;.\quad\quad\quad\quad\;\;
\label{eq10}
\end{eqnarray}
In the ${U(1)}_{L^{}_{\mu} - L^{}_{\tau}}$ model, $Z^{}_{\mu\tau}$ has a light mass and no couplings to the SM quarks and the first generation leptons, so it can only decay to neutrinos. The couplings of $Z^{}_{\mu\tau}$ with neutrinos are expressed as
\begin{eqnarray}
g^{}_{Z^{}_{\mu\tau}v^{}_{\mu}v^{}_{\mu}}= \frac{{M}^{}_{Z^{}_{\mu\tau}}} { v^{}_{\mu\tau}}\;,\hspace{1cm} g^{}_{Z^{}_{\mu\tau}v^{}_{\tau}v^{}_{\tau}} = -\frac{{M}^{}_{Z^{}_{\mu\tau}}} { v^{}_{\mu\tau}}\;.
\end{eqnarray}
 Taking no account of the neutrino masses, the expression form of the total decay width of $Z^{}_{\mu\tau}$ is given by
\begin{eqnarray}
\Gamma^{}_{Z^{}_{\mu\tau}}=\frac{g^{2}_{\mu\tau}M^{}_{Z^{}_{\mu\tau}}}{12\pi}\;,\qquad
\end{eqnarray}
where we have also ignored the neutrino mixing.

To produce the appropriate neutrino mass and explain the muon $(g-2)$ anomaly, Ref.\cite{Nomura:2018yej} has show the favored regions of the gauge coupling  $g^{}_{\mu\tau}$ and the $Z^{}_{\mu\tau}$ mass, which are summarized as
\begin{eqnarray}
g^{}_{\mu\tau}\simeq[2\times10^{-4}\;,\;2\times10^{-3}]\;,\hspace{0.7cm}
{M}_{Z^{}_{\mu\tau}}\simeq[5\;,\;210]~{\rm MeV}\;.
\label{eq11}
\end{eqnarray}
According to Eq.~(\ref{eq11}), the range of $v^{}_{\mu\tau}$ is given by
\begin{eqnarray}
v^{}_{\mu\tau} = \frac{{M}_{Z^{}_{\mu\tau}}} { g^{}_{\mu\tau}}\simeq [10\;,\;1000 ]~{\rm GeV}\;,\quad
\end{eqnarray}
which indicates that, after the spontaneous symmetry breaking, the new scalar $h_{2}$ obtains  mass as the same order with $v^{}_{\mu\tau}$.

From above discussions we can see that, besides decaying to SM particles, the SM-like Higgs boson $h^{}_{1}$ has extra decay modes  $Z^{}_{\mu\tau}Z^{}_{\mu\tau}$ and $\varphi^{\dagger}_{DM}\varphi^{}_{DM}$ for $M^{}_{DM}<M^{}_{h^{}_{1}}$. The expressions of these decay channels are
\begin{eqnarray}
\Gamma(h^{}_{1}\to Z^{}_{\mu\tau}Z^{}_{\mu\tau})=\frac{g^{2}_{Z^{}_{\mu\tau}Z^{}_{\mu\tau}h^{}_{1}}({M}^{4}_{h^{}_{1}}-{4M}^{2}_{Z^{}_{\mu\tau}}{M}^{2}_{h^{}_{1}}+12{M}^{4}_{Z^{}_{\mu\tau}})\sqrt{{M}^{2}_{h^{}_{1}}-{4M}^{2}_{Z^{}_{\mu\tau}}}}{128 \pi{M}^{2}_{h^{}_{1}}{M}^{4}_{Z^{}_{\mu\tau}}}\;,
\end{eqnarray}
\begin{eqnarray}
\Gamma(h^{}_{1}\to \varphi^{\dagger}_{DM}\varphi^{}_{DM})=\frac{g^{2}_{\varphi^{\dagger}_{DM}\varphi^{}_{DM}h^{}_{1}}}{32 \pi{M}^{}_{h^{}_{1}}}\sqrt{1-\frac{{4M}^{2}_{DM}}{{M}^{2}_{h^{}_{1}}}}\;,
\end{eqnarray}
where the $M^{}_{DM}$ is Dark Matter mass. The total decay width of $h^{}_{1}$ can be written as
\begin{eqnarray}
\Gamma(h^{}_{1})=\cos^{2}_{}\alpha\Gamma^{}_{SM}+\Gamma(h^{}_{1}\to Z^{}_{\mu\tau}Z^{}_{\mu\tau})+\Gamma(h^{}_{1}\to \varphi^{\dagger}_{DM}\varphi^{}_{DM})\;,
\end{eqnarray}
where $\Gamma^{}_{SM}$ is the total width of the Higgs boson in the SM. In the ${U(1)}_{L^{}_{\mu} - L^{}_{\tau}}$ model, the decays of $\varphi^{}_{DM}$ and $Z^{}_{\mu\tau}$ are invisible, the branching ratio of the invisible decays is given by
\begin{eqnarray}
BR(h^{}_{1}\to {\rm invisibles})=\frac{\Gamma(h^{}_{1}\to Z^{}_{\mu\tau}Z^{}_{\mu\tau})+\Gamma(h^{}_{1}\to \varphi^{\dagger}_{DM}\varphi^{}_{DM})}{\Gamma(h^{}_{1})}\;.
\end{eqnarray}
As we can see from Eq.~(\ref{eq9}) and Eq.~(\ref{eq10}), the couplings of ${\varphi}^{}_{DM}$ with scalar bosons $h_{1}$ and $h_{2}$ depend on the parameters $\lambda_{Dh}$ and $\lambda_{DH}$. In this work, we take $\lambda_{Dh}=9.8\times10^{-13}$ and $\lambda_{DH}=1.3\times10^{-11}$~\cite{Biswas:2016yjr}, which makes the $\Gamma(h^{}_{1}\to \varphi^{\dagger}_{DM}\varphi^{}_{DM})$ and $\Gamma(h^{}_{2}\to \varphi^{\dagger}_{DM}\varphi^{}_{DM})$ so feeble that they can be ignored.
Using the constraint on the branching ratio of the Higgs invisible decay, $\rm BR_{invis}\leq 0.24$ at $95\%$ C. L. from the LHC data~\cite{BSM52}, the sine of scalar mixing angle $\sin\alpha$ must be satisfied
$	\sin\alpha\leq 0.3$. Then, for the factor $ \chi $, there is
	\begin{eqnarray}
	\chi =\frac{\alpha}{v_{\mu\tau}}\leq2.2\times10^{-4} ~\rm{GeV^{-1}}\;.
	\end{eqnarray}	
	
To summarize, in the ${U(1)}_{L^{}_{\mu} - L^{}_{\tau}}$ model, three new free parameters are introduced, which are the new gauge coupling constant $g^{}_{\mu\tau}$, the $Z^{}_{\mu\tau}$ mass $M_{Z_{\mu\tau}}$ and the scalar mixing angle $\alpha$, respectively. In the following, we will focus our attention on the phenomenology of the new particles $h_{2}$ and $Z^{}_{\mu\tau}$ at e-p colliders in the above allowed parameter space.

\section*{\uppercase\expandafter{\romannumeral3}. Decays and Productions of the  Scalar $h^{}_{2}$ }
\textbf{ 3.1. Decays of the scalar $h^{}_{2}$ }
\\
\begin{figure}[!ht]
	\centering
	\subfigure{
		\begin{minipage}[b]{0.7\textwidth}
			\includegraphics[width=1\textwidth]{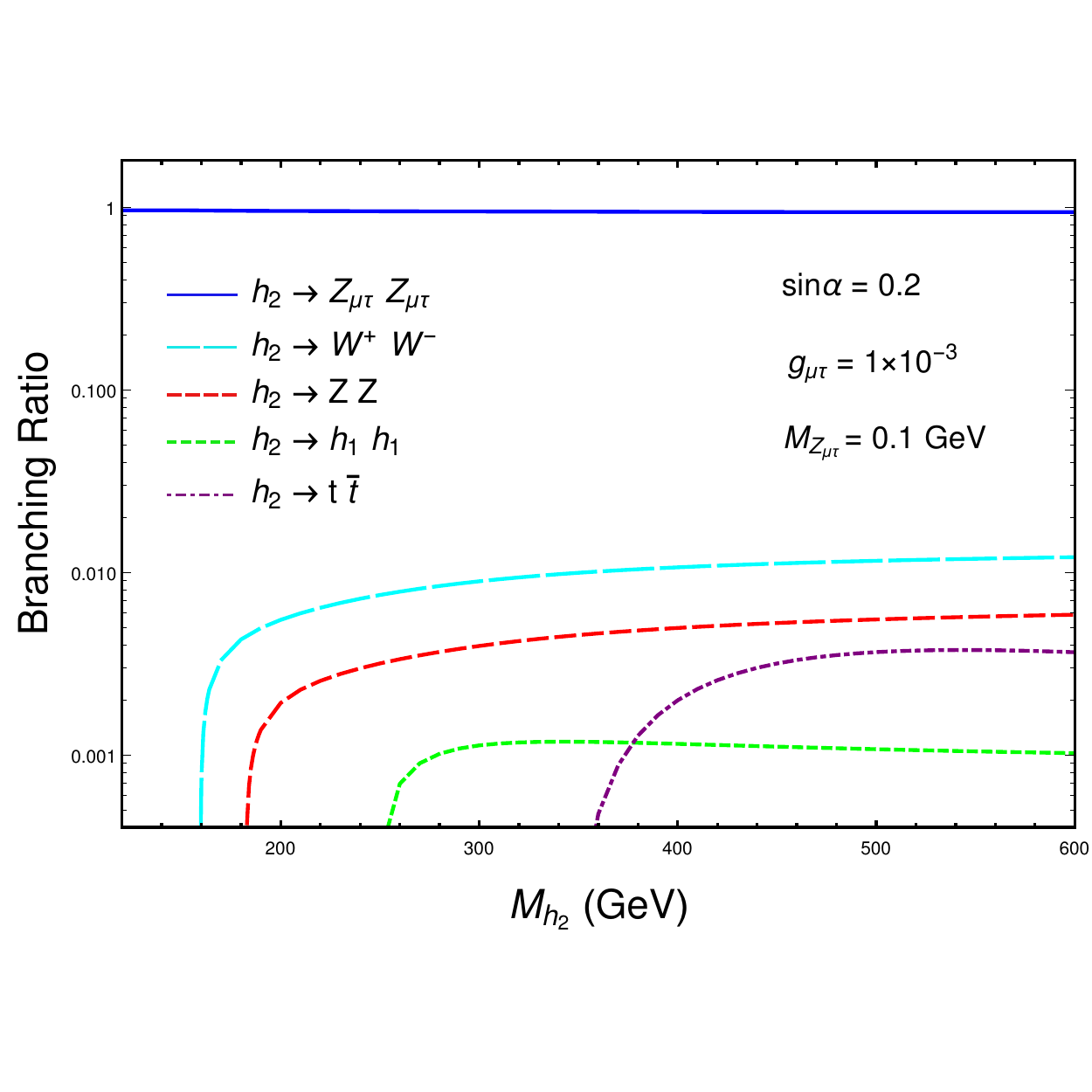}
		\end{minipage}
0	}
	\caption{\footnotesize The branching ratios for the main decay modes of the scalar $h_{2}$ as functions of $M_{h_{2}}$ for the fixed values sin$\alpha$ = 0.2, $M_{Z_{\mu\tau}}=0.1$ GeV and  $g^{}_{\mu\tau}=1\times10^{-3}$.}
	\label{fig1}
\end{figure}

In the ${U(1)}_{L^{}_{\mu}-L^{}_{\tau}}$  model, the scalar $h_{2}$ can not only decay to the SM particles but also decay to the new particles $Z_{\mu\tau}$ and $\varphi^{}_{DM}$. Here, we give the decay width expressions of its several major decay modes.  The expression form of the decay width for the decay channel $h_{2}\to Z_{\mu\tau}Z_{\mu\tau}$ is given by
\begin{eqnarray}
\Gamma(h^{}_{2}\to Z^{}_{\mu\tau}Z^{}_{\mu\tau})=\frac{g^{2}_{Z^{}_{\mu\tau}Z^{}_{\mu\tau}h^{}_{2}}({M}^{4}_{h^{}_{2}}-{4M}^{2}_{Z^{}_{\mu\tau}}{M}^{2}_{h^{}_{2}}+12{M}^{4}_{Z^{}_{\mu\tau}})\sqrt{{M}^{2}_{h^{}_{2}}-{4M}^{2}_{Z^{}_{\mu\tau}}}}{128\pi{M}^{2}_{h^{}_{2}}{M}^{4}_{Z^{}_{\mu\tau}}}\;.
\end{eqnarray}
Under the assumption $\frac{{M}^{}_{Z^{}_{\mu\tau}}}{{M}^{}_{h^{}_{1}}}\to0$, we can obtain the following form
\begin{eqnarray}
\Gamma(h^{}_{2}\to Z^{}_{\mu\tau}Z^{}_{\mu\tau})=\frac{{M}^{3}_{h^{}_{2}}\cos^{2}\alpha}{32 \pi v^{2}_{\mu\tau}}\;.
\end{eqnarray}

The decay width of the channel $h_{2}\to \varphi^{\dagger}_{DM}\varphi^{}_{DM}$ is
\begin{eqnarray}
\Gamma(h^{}_{2}\to \varphi^{\dagger}_{DM}\varphi^{}_{DM})=\frac{g^{2}_{\varphi^{\dagger}_{DM}\varphi^{}_{DM}h^{}_{2}}}{16 \pi{M}^{}_{h^{}_{2}}}\sqrt{1-\frac{{4M}^{2}_{DM}}{{M}^{2}_{h^{}_{2}}}}\;.
\end{eqnarray}
As we already mentioned, the value of $\Gamma(h^{}_{2}\to \varphi^{\dagger}_{DM}\varphi^{}_{DM})$ is small enough and we will neglect it.

The width of $h_{2}$ decaying to vector bosons is given as
\begin{eqnarray}
\Gamma(h^{}_{2}\to VV)=\frac{g^{2}_{VVh^{}_{2}}({M}^{4}_{h^{}_{2}}-{4M}^{2}_{V}{M}^{2}_{h^{}_{2}}+12{M}^{4}_{V})\sqrt{{M}^{2}_{h^{}_{2}}-{4M}^{2}_{V}}}{64 \pi S_{V}{M}^{2}_{h^{}_{2}}{M}^{4}_{V}}\;,
\end{eqnarray}
where $S_{V}$ represents the statistical factor. Its value equals to 1 for $W^{\pm}$ boson and 2 for $Z$ boson.

The width for the decay process $h^{}_{2}\to h^{}_{1}h^{}_{1}$ can be written as
\begin{eqnarray}
\Gamma(h^{}_{2}\to h^{}_{1}h^{}_{1})=\frac{g^{2}_{h^{}_{1}h^{}_{1}h^{}_{2}}\sqrt{{M}^{2}_{h^{}_{2}}-{4M}^{2}_{h^{}_{1}}}}{32 \pi{M}^{2}_{h^{}_{2}}}\;.
\end{eqnarray}
From Eq.~(\ref{eq10}) one can see that the coupling constant $g_{h_{1}h_{1}h_{2}}$ depends on the couplings $\lambda^{}_{h}$,  $\lambda^{}_{H}$ and $\lambda^{}_{hH}$, which are expressed as
\begin{eqnarray}
\lambda^{}_{H}\! \! \! &=&\! \! \! \frac{{{M}^2_{h^{}_{2}}}+{{M}^2_{h^{}_{1}}}-({{M}^2_{h^{}_{1}}}-{{M}^2_{h^{}_{2}}})\cos2\alpha}{4v^{2}_{\mu\tau}}\;,\quad\nonumber \\
\lambda^{}_{h}\! \! \! &=&\! \! \! \frac{{{M}^2_{h^{}_{2}}}+{{M}^2_{h^{}_{1}}}-({{M}^2_{h^{}_{2}}}-{{M}^2_{h^{}_{1}}})\cos2\alpha}{4v^{2}}\;,\quad\nonumber \\
\lambda^{}_{hH}\! \! \! &=&\! \! \! -\frac{({{M}^2_{h^{}_{2}}}-{{M}^2_{h^{}_{1}}})\cos\alpha\sin\alpha}{vv^{}_{\mu\tau}}\;.
\end{eqnarray}
When we take $\sin\alpha=0.2$ and $v_{\mu\tau}=\frac{M_{Z_{\mu\tau}}}{g_{\mu\tau}}=100$ GeV, the decay width $\Gamma(h^{}_{2}\to h^{}_{1}h^{}_{1})$ only depends on $M_{h_{2}}$.

The width of $h_{2}$ decaying to the SM fermion pair is given as
\begin{eqnarray}
\Gamma(h^{}_{2}\to f\bar{f})=\frac{n_{c}g_{f\bar{f}h_{2}}({M}^{2}_{h^{}_{2}}-{4M}^{2}_{f})^{\frac{3}{2}}}{8 \pi{M}^{2}_{h^{}_{2}}}\;,
\end{eqnarray}
where the color charge $n_{c}=1$ for leptons and 3 for quarks.
Fig.~\ref{fig1} shows the  branching ratios for the main decay modes of the scalar $h_{2}$ as functions of the mass parameter $M_{h_{2}}$ for the fixed values sin$\alpha$ = 0.2, $M_{Z_{\mu\tau}}=0.1$ GeV and $g^{}_{\mu\tau}=1\times10^{-3}$, where the curves from high to low correspond the
$Z_{\mu\tau}Z_{\mu\tau}$ decay modes, the $W^{+}W^{-}/ZZ$ decay modes, di-higgs decay mode and di-top decay mode, respectively. One can see from this figure that the value of the branching ratio $BR(h_{2}\to Z_{\mu\tau}Z_{\mu\tau})$ is about $98\%$ and only is $2\%$ for the rest decay channels. Certainly, the values of these  branching ratios would vary as the values of the parameters sin$\alpha$ and $g_{\mu\tau}$ changing. However, in the allowed parameter space of the ${U(1)}_{L^{}_{\mu}-L^{}_{\tau}}$  model, the decay
process $h_{2}\to Z_{\mu\tau}Z_{\mu\tau}$ is the main decay channel of the scalar $h_{2}$. \\

\textbf{ 3.2. Productions of the scalar $h^{}_{2}$ }\\

\begin{figure}[!ht]
	\centering
	\subfigure[]{
		\begin{minipage}[b]{0.27\textwidth}
			\includegraphics[width=1\textwidth]{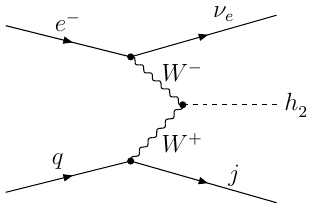}
		\end{minipage}
	}
	\quad\quad\quad\quad\quad
	\subfigure[]{
		\begin{minipage}[b]{0.28\textwidth}
			\includegraphics[width=1\textwidth]{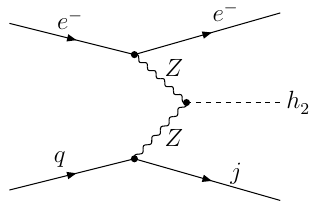}
		\end{minipage}
	}
	\caption{\footnotesize The Feynman diagrams for the scalar $h_{2}$ productions at e-p colliders (Left (a): CC production channel, Right (b): NC production channel). }
	\label{fig2}
\end{figure}

\begin{figure}[!ht]
	\centering
	\subfigure[]{
		\begin{minipage}[b]{0.47\textwidth}
			\includegraphics[width=1\textwidth]{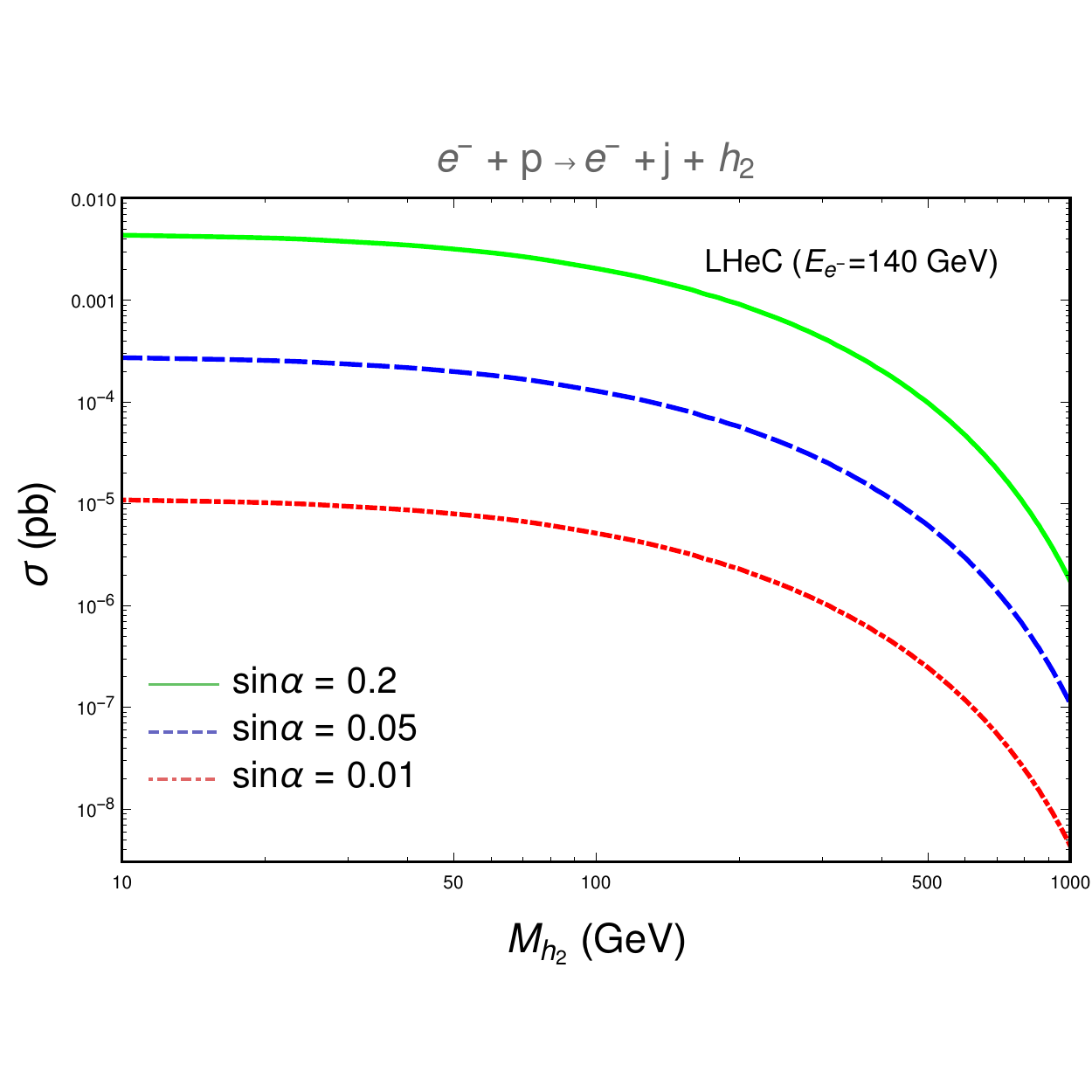}
		\end{minipage}
	}
	\subfigure[]{
		\begin{minipage}[b]{0.473\textwidth}
			\includegraphics[width=1\textwidth]{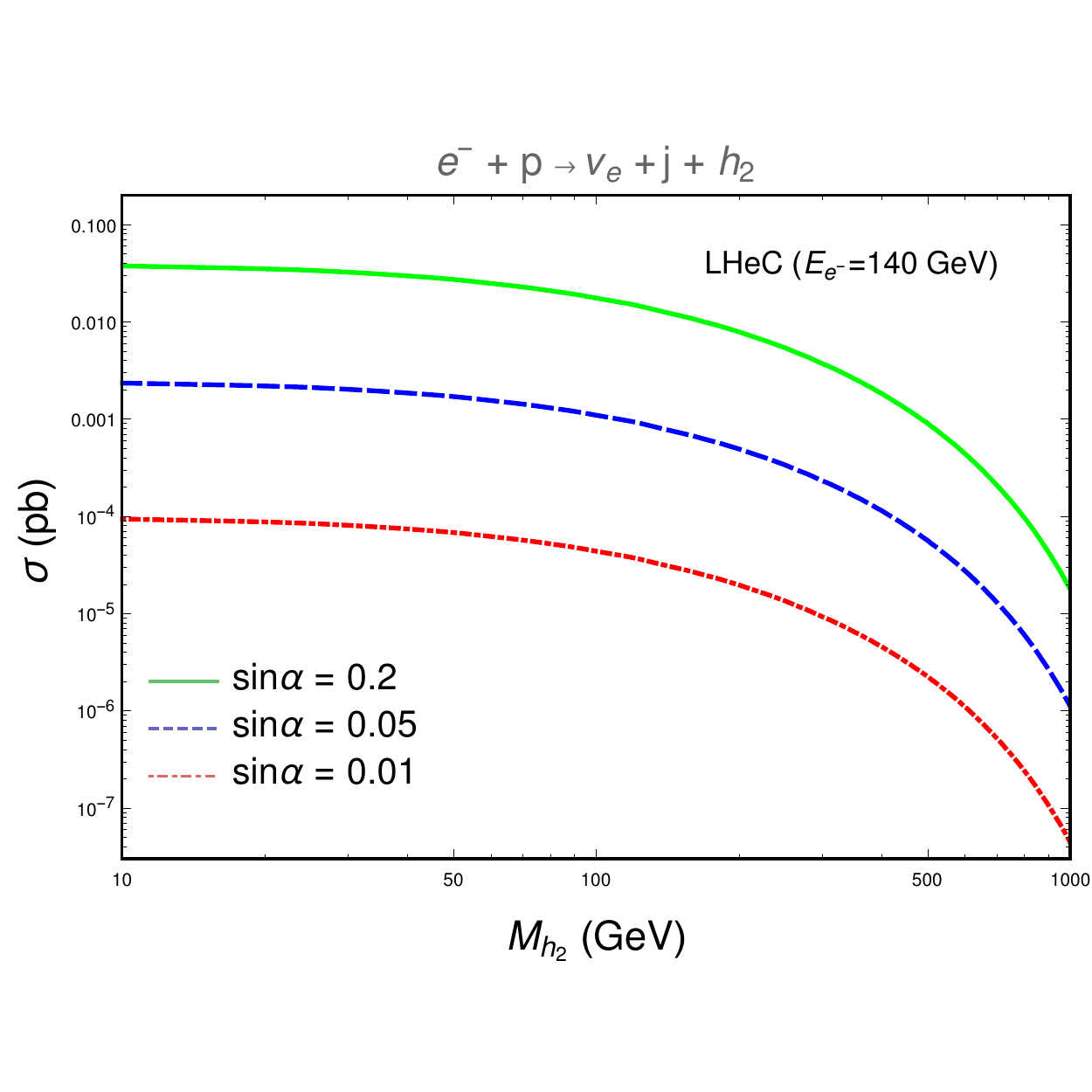}
		\end{minipage}
	}
\caption{\footnotesize The production cross sections of the processes $e^{-}p\to{e^{-}jh_{2}}$ and $e^{-}p\to{\nu_{e}j h_{2}}$ as functions of the mass parameter $M_{h_{2}}$ for $\sin\alpha=0.2$ (solid),~$0.05$ (dashed),~$0.01$ (dotted) and the beam polarization P($e^{-}$)= -0.8 at the LHeC.}
\label{fig3}
\end{figure}

Like the SM Higgs boson, the additional scalar $h^{}_{2}$ in the ${U(1)}_{L^{}_{\mu}-L^{}_{\tau}}$  model is produced via two major channels: the charged current (CC) production channel via $W^{+}W^{-}$ fusion and the neutral current (NC) production channel via $ZZ$ fusion~\cite{Tang:2015uha,Han:2009pe} at e-p colliders. Fig.~\ref{fig2} gives the corresponding Feynman diagrams for the $h_{2}$ production via CC production channel and NC production channel at e-p colliders, respectively.
Then, employing Madgraph5/aMC@NLO~\cite{Alwall:2014hca}, we calculate the production cross sections of the  processes  $e^{-}p\to{\nu_{e}j h_{2}}$ and $e^{-}p\to{e^{-}jh_{2}}$ as functions
of $M_{Z_{\mu\tau}}$ at the LHeC. It is well known that polarization of the initial state electron can affect the production cross sections. Our numerical results show that the beam polarization P($e^{-}$)= -0.8 can maximize the cross sections. Therefore, we will take $P(e^{-})=-0.8$ in following numerical calculation. Since $g_{\mu\tau}$ does not affect the cross sections of $h_{2}$ production via the CC and NC processes, we do not consider it here. In Figs.~\ref{fig3} (a) and ~\ref{fig3} (b), the curves show the cross sections of the $e^{-}p\to{e^{-}j h_{2}}$ and $e^{-}p\to{\nu_{e}jh_{2}}$ processes with $E_{e^{-}}=140$ GeV and different values of mixing angle $\sin\alpha=0.2$ (solid),~$0.05$ (dashed) and~$0.01$ (dotted). One can see from these figures that the values of the production cross section $\sigma$ decrease as the $h_{2}$ mass increases. For the $e^{-}p\to{e^{-}j h_{2}}$ process and 10 GeV $ \leq{M}_{h_{2}}\leq $ 1000 GeV, its values are in the ranges of $1.73~\rm{pb}\times10^{-6}\leq\sigma\leq4.33\times10^{-3}$ pb~(solid), $1.08\times10^{-7}~\rm{pb}\leq\sigma\leq2.71\times10^{-4}$ pb ~(dashed) and $4.33\times10^{-9}~\rm{pb}\leq\sigma\leq1.08\times10^{-5}$ pb~(dotted), respectively. For the $e^{-}p\to{\nu_{e}jh_{2}}$ process and 10 GeV $ \leq{M}_{h_{2}}\leq $ 1000 GeV, its values are in the ranges of $1.77\times10^{-5}~\rm{pb}\leq\sigma\leq3.73\times10^{-2}$ pb~(solid), $1.11\times10^{-6}~\rm{pb}\leq\sigma\leq2.33\times10^{-3}$ pb ~(dashed) and $4.43\times10^{-8}~\rm{pb}\leq\sigma\leq9.32\times10^{-5}$ pb~(dotted) respectively. It is worth mentioning that the cross section of the $e^{-}p\to{\nu_{e}j h_{2}}$ process is larger than that of the $e^{-}p\to{e^{-}j h_{2}}$ process by about one order of magnitude.

\section*{\uppercase\expandafter{\romannumeral4}. Productions of the New Gauge Boson $Z_{\mu\tau}$ }

Now, we turn our attention to the new gauge boson $Z_{\mu\tau}$. As mentioned in the previous section, $Z^{}_{\mu\tau}$ can not establish couplings with all the SM quarks and the first generation leptons, making it very difficult to be produced directly. So it is a attractive scheme to obtain $Z^{}_{\mu\tau}$ by considering its indirect production. Similar with the new scalar $h_{2}$, besides decaying to the SM particles, the SM-like Higgs boson $h^{}_{1}$ can also decay  to a pair of $Z^{}_{\mu\tau}$. Eq.~(15) has given the expression form of the decay width $\Gamma(h^{}_{1}\to Z^{}_{\mu\tau}Z^{}_{\mu\tau})$, which can be simplified to
\begin{eqnarray}
\Gamma(h^{}_{1}\to Z^{}_{\mu\tau}Z^{}_{\mu\tau})=\frac{{M}^{3}_{h^{}_{1}}\sin^{2}\alpha}{32 \pi v^{2}_{\mu\tau}}\;.
\label{eq22}
\end{eqnarray}
From above equation we can see that the production rate of the $Z^{}_{\mu\tau}$ pair from $h^{}_{1}$ decaying is actually determined by the factor $\chi^{2} \simeq\sin^{2}\alpha/v^{2}_{\mu\tau}$. So, in this work, all the results for the $Z^{}_{\mu\tau}$ production via $h^{}_{1}$ decaying  can be expressed as functions of the factor $\chi$. Next, we will consider its indirect productions via the decays of $h^{}_{2}$ and $h^{}_{1}$, respectively.

\begin{figure}[!ht]
	\centering
	\subfigure[]{
		\begin{minipage}[b]{0.3\textwidth}
			\includegraphics[width=1\textwidth]{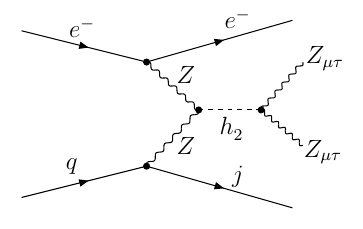}
		\end{minipage}
	}
\quad\quad\quad\quad\quad
\subfigure[]{
	\begin{minipage}[b]{0.31\textwidth}
		\includegraphics[width=1\textwidth]{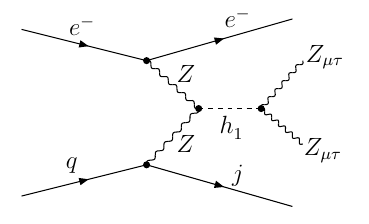}
	\end{minipage}
}
	\caption{\footnotesize The leading order Feynman diagrams of the $Z^{}_{\mu\tau}$  productions by the decays of $h^{}_{1}$ and $h^{}_{2}$ via the NC production channels at e-p colliders.}
	\label{fig4}
\end{figure}

As can be seen from  Fig.~\ref{fig3}, the CC production of scalar $h_{2}$ has larger cross section than that for its NC production.  However, as mentioned earlier, $Z^{}_{\mu\tau}$ can only decay to neutrinos in the ${U(1)}_{L^{}_{\mu}-L^{}_{\tau}}$  model. So the final states of the $e^{-}p\to{\nu_{e} j h^{}_{2}}(h^{}_{2}\to{Z^{}_{\mu\tau}Z^{}_{\mu\tau}})$ and $e^{-}p\to{\nu_{e} j h^{}_{1}}(h^{}_{1}\to{Z^{}_{\mu\tau}Z^{}_{\mu\tau}})$ processes would be jets and missing energy, which are difficult to be distinguished from the deeply inelastic scattering (DIS) backgrounds. Moreover, lack of kinematic handles in the final state makes it extremely difficult to filter signal from many backgrounds. Therefore, in this work we will focus on the NC production channels $e^{-}p\to{e^{-}jh_{1}(\to{Z^{}_{\mu\tau}Z^{}_{\mu\tau}})\to}~e^{-}j+\cancel{E}^{}_{T}$ and $e^{-}p\to{e^{-}jh_{2}(\to{Z^{}_{\mu\tau}Z^{}_{\mu\tau}})\to}~e^{-}j+\cancel{E}^{}_{T}$ to study the feasibility of detecting $h_{2}$ and $Z^{}_{\mu\tau}$ .  In Fig.~\ref{fig4}, we show the leading order Feynman diagrams of the $Z^{}_{\mu\tau}$  productions by the decays of $h^{}_{1}$ and $h^{}_{2}$ at e-p colliders.
\begin{figure}[!ht]
	\centering
%	\subfigure[]{
%		\begin{minipage}[b]{0.465\textwidth}
%			\includegraphics[width=1\textwidth]{figure8.pdf}
%		\end{minipage}
%	}
	\subfigure{
		\begin{minipage}[b]{0.7\textwidth}
			\includegraphics[width=1\textwidth]{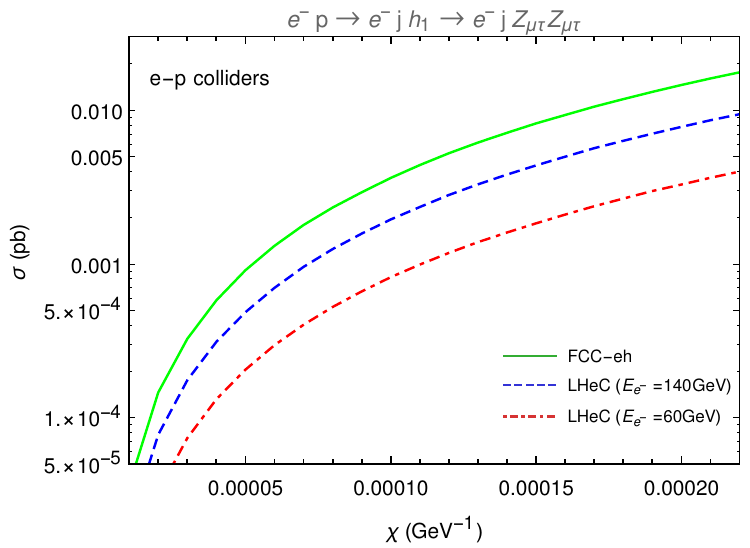}
		\end{minipage}
	}
	\caption{\footnotesize The production cross sections of the process $e^{-}p\to{e^{-}jh_{1}\to}~e^{-}jZ^{}_{\mu\tau}Z^{}_{\mu\tau}$ as functions of $\chi$ at e-p colliders.}
	\label{fig5}
\end{figure}

Employing the Madgraph5/aMC@NLO~\cite{Alwall:2014hca}, we calculate the cross sections of the $Z^{}_{\mu\tau}$ production processes  $e^{-}p\to{e^{-}jh_{2}\to}~e^{-}jZ^{}_{\mu\tau}Z^{}_{\mu\tau}$ and $e^{-}p\to{e^{-}jh_{1}\to}~e^{-}jZ^{}_{\mu\tau}Z^{}_{\mu\tau}$. Considering the favored region of the parameter space to resolve $(g-2)_{\mu}$ discrepancy, $g_{\mu\tau}$ is fixed to $g_{\mu\tau}=1\times10^{-3}$ for reference. Due to $BR(h_{2}\to Z^{}_{\mu\tau}Z^{}_{\mu\tau})\sim1$, the cross section of the $Z^{}_{\mu\tau}$ production via $h_{2}$ decaying as a function of the mass $M_{h_{2}}$ for $\sin\alpha=0.2$,~$0.05$ and~$0.01$ at LHeC with $E_{e^{-}}=140$ GeV is essentially the same as Fig.~\ref{fig3} (a). So we don't show it again. Fig.~\ref{fig5} shows the cross sections of the $Z^{}_{\mu\tau}$ production by $h_{1}$ decaying as functions of the factor $\chi$ at e-p colliders, where the different curves show the $Z_{\mu\tau}$ production cross sections at different colliders: FCC-eh (solid),~LHeC with $E_{e^{-}}=140~\rm{GeV}$ (dashed) and~~LHeC with $E_{e^{-}}=60~\rm{GeV}$ (dotted). From Fig.~\ref{fig5}, we can see that, for $1\times10^{-5}$ GeV $ \leq\chi\leq $ $2\times10^{-4}$ GeV, the values of the $Z_{\mu\tau}$ production cross sections are in the ranges $3.61\times10^{-5}~\rm{pb}\leq\sigma\leq1.75\times10^{-2}$ pb~(solid), $1.94\times10^{-5}~\rm{pb}\leq\sigma\leq9.37\times10^{-3}$ pb ~(dashed) and $8.16\times10^{-6}~\rm{pb}\leq\sigma\leq3.97\times10^{-3}$ pb~(dotted).
\section*{\uppercase\expandafter{\romannumeral5}. Signatures of the New Particles $h_{2}$ and $Z_{\mu\tau}$ at e-p Colliders}

In this section, we analyze the observation potential by performing a Monte Carlo simulation of the signal and background events and explore the observability of the additional scalar $h_{2}$ and the new gauge boson $Z^{}_{\mu\tau}$ at e-p colliders with the integrated luminosity of 1 $\rm{ab^{-1}}$. On one hand, we will explore the observability of $h_{2}$ as well as $Z^{}_{\mu\tau}$ via the processes $e^{-}p\to{e^{-}jh_{2}\to}~e^{-}jZ^{}_{\mu\tau}Z^{}_{\mu\tau}$ and $e^{-}p\to{e^{-}jh_{1}\to}~e^{-}jZ^{}_{\mu\tau}Z^{}_{\mu\tau}$. For the second process, we mainly use it to explore  the signature of $Z^{}_{\mu\tau}$. On the other hand, for more comprehensive, we  will also analysis the possibility of detecting  $h_{2}$ through CC production channel followed by $h_{2}$ decaying into vector boson pair ($ZZ$).

We use Madgraph5/aMC@NLO~\cite{Alwall:2014hca} to calculate the relevant production cross sections and generate the signal and background events, where the UFO format of the  ${U(1)}_{L^{}_{\mu} - L^{}_{\tau}}$ model has been obtained by using  FeynRules~\cite{Christensen:2008py}. Moreover, the parton distribution function (PDF), NNPDF2.3~\cite{Ball:2013hta}, is used  at leading order and Pythia-pgs~\cite{Sjostrand:2006za} is employed for parton showering, hadronization and fast detector simulation. Finally, MadAnalysis5~\cite{Conte:2012fm} is applied for data analysis and plotting. All of the SM input parameters are taken from Particle Data Group (PDG)~\cite{Tanabashi:2018oca}.\\

\textbf{ A. $h_{2}\to Z_{\mu\tau}Z_{\mu\tau}$ and  $h_{1}\to Z_{\mu\tau}Z_{\mu\tau}$ Channels}\\

In this subsection, we take both $h_{2}$ and $h_{1}$ productions at e-p colliders through NC production channels followed by
$h^{}_{2}\to{Z^{}_{\mu\tau}Z^{}_{\mu\tau}}$ and $h^{}_{1}\to{Z^{}_{\mu\tau}Z^{}_{\mu\tau}}$ as our signals, signal-1 and signal-2, respectively. Since $Z_{\mu\tau}$ has invisible final state in the detector, these two processes provide the same final state that includes one electron, one jet and a large missing transverse energy $\cancel{E}^{}_{T}$
\begin{eqnarray}
e^{-}+p\to{e^{-}+j}+{h^{}_{2}(\to{Z^{}_{\mu\tau}Z^{}_{\mu\tau}})\to}~e^{-}+j+\cancel{E}^{}_{T}\;,~(\rm{signal-1})
\end{eqnarray}
\begin{eqnarray}
e^{-}+p\to{e^{-}+j}+{h^{}_{1}(\to{Z^{}_{\mu\tau}Z^{}_{\mu\tau}})\to}~e^{-}+j+\cancel{E}^{}_{T}\;, ~(\rm{signal-2})
\end{eqnarray}
in which $\cancel{E}^{}_{T}$ comes from $Z^{}_{\mu\tau}\to\nu\bar{\nu}$. For $\sin\alpha=0.2$ and $g_{\mu\tau}=1\times10^{-3}$, the values of the cross section for the signal-1 are $4.179\times10^{-4}$ pb ($1.046\times10^{-4}$ pb) for $E_{e^{-}}=140~(60)$ GeV at the LHeC and $9.251\times10^{-4}$ pb at the FCC-eh . While the values of the cross section for the signal-2 are $1.898\times10^{-3}$ pb ($5.441\times10^{-4}$ pb) for $E_{e^{-}}=140~(60)$ GeV at the LHeC and $2.412\times10^{-3}$ pb at the FCC-eh for $\chi=9\times10^{-5}~\rm{GeV^{-1}}$.

For the signal $ e^{-}j \cancel{E}_{T} $, the leading irreducible SM backgrounds can be classified into two general categories. The first category has a final state $e^{-}j\nu_{e}\bar{\nu}_{e}$ which comes from the following two processes
\begin{eqnarray}
e^{-}+p\to W^{-}(\to e^{-}\bar{\nu}_{e})+j+\nu_{e}\to e^{-}+j+\cancel{E}^{}_{T}~(e^{-}j\nu_{e}\bar{\nu}_{e})\;,
\end{eqnarray}
\begin{eqnarray}
\!\!\!\!\!\!\!e^{-}+p\to Z(\to\nu_{e}\bar{\nu}_{e})+j+e^{-}\to e^{-}+j+\cancel{E}^{}_{T}~(e^{-}j\nu_{e}\bar{\nu}_{e})\;.
\end{eqnarray}
The total cross section for this kind of irreducible backgrounds is 0.4334 pb (0.205 pb) for $E_{e^{-}}=140~(60)$ GeV at the LHeC and 0.8116 pb at the FCC-eh, which will severely pollute the physical signal.

The second category has a final state $e^{-}j\nu_{\mu,\tau}\bar{\nu}_{\mu,\tau}$
\begin{eqnarray}
\!\!\!\!\!\!\!e^{-}+p\to Z(\to\nu_{\mu,\tau}\bar{\nu}_{\mu,\tau})+j+e^{-}\to e^{-}+j+\cancel{E}^{}_{T}~(e^{-}j\nu_{\mu,\tau}\bar{\nu}_{\mu,\tau})\;.
\end{eqnarray}
Its production cross section is 0.05685 pb (0.03422 pb) for $E_{e^{-}}=140~(60)$ GeV at the LHeC and 0.1052 pb at the FCC-eh.  Besides, more remarkably, the photoproduction of the state $W+j$, which has a larger cross section, is also an irreducible SM background if the $W$ boson decays to an electron and neutrino. But it can be negligible after all selection cuts, because of its unique kinematic features.

There are also some reducible backgrounds which come from various sources. The most threatening reducible backgrounds result from the production of $\tau$ in the final state. One is
\begin{eqnarray}
e^{-}+p\to e^{-}+j+\tau^{+}+\nu_{\tau} ~ (e^{-}j\tau^{+}\nu_{\tau})\;.
\label{eq28}
\end{eqnarray}
Its cross section is 0.264 pb (0.1331 pb) for $E_{e^{-}}=140~(60)$ GeV at the LHeC and 0.3957 pb at the FCC-eh. The other one is
\begin{eqnarray}
e^{-}+p\to e^{-}+j+\tau^{-}+\bar{\nu}_{\tau}~(e^{-}j\tau^{-}\bar{\nu}_{\tau})\;.
\label{eq29}
\end{eqnarray}
Its cross section is 0.2816 pb (0.1362 pb) for $E_{e^{-}}=140~(60)$ GeV at the LHeC and 0.5015 pb at the FCC-eh. The main reasons why the above two processes (Eq.~(\ref{eq28}) and Eq.~(\ref{eq29})) can be viewed as reducible backgrounds are: (I) The $\tau$-jets may be misidentified as hadronic jets. (II) The detection of hadronic decay products of $\tau$ cannot be expected to be fully efficient due to the products being too soft, which will lead to generation of the missing energy $\cancel{E}^{}_{T}$. Furthermore, we can even consider the case (II) as a source of partial irreducible background. The $e^{-}p\to \nu_{e}j\tau^{+}\nu_{\tau}$ and $e^{-}p\to \nu_{e}j\tau^{-}\bar{\nu}_{\tau}$ processes are reducible backgrounds in which the $\tau$ decays to an electron. Fortunately, we could suppress them to an insignificant order because of the totally different kinematic distribution of the final electron. Some other reducible backgrounds  are $e$+multijet productions in which the $\cancel{E}^{}_{T}$ comes from jet's mismeasurement and $jj\nu$ production in which one jet is misidentified as an electron. In this work, we do not simulate both of them because their contributions can be negligible after all selection cuts.
The signal and background events are generated with following basic cuts~\cite{Tang:2015uha} in Madgraph5/aMC@NLO~\cite{Alwall:2014hca}\\

\qquad \qquad \qquad \textbullet ~lepton transverse momentum ~~$ p^{}
_{T}(l^{\pm})> 5$  GeV,\\

\qquad \qquad \qquad \textbullet ~jet transverse momentum ~~$ p^{}
_{T}(j)> 20$  GeV,\\

\qquad \qquad \qquad \textbullet ~~~lepton pseudorapidity in the range $\vert \eta(l^{\pm})\vert<5$\;,\\

\qquad \qquad \qquad \textbullet ~~~jet pseudorapidity in the range $\vert \eta(j)\vert<5$\;,\\

\qquad \qquad \qquad \textbullet~~~angular separation between jet and lepton $\Delta R(jl^{\pm})>0.4$\;,\\
\\
where $\eta=1/2\ln(\tan\theta)$ is the pseudorapidity, where $\theta$  indicates the scattering angle in the laboratory frame. $\Delta R = \sqrt{(\Delta\phi)^{2}+(\Delta\eta)^{2}}$ is the particle separation, where $\Delta\phi$ and $\Delta\eta$ represent the rapidity gap
and the azimuthal angle gap between the particle pair, respectively.
\begin{figure}[htbp]
	\subfigure[]{
		\begin{minipage}[b]{0.45\textwidth}
			\includegraphics[width=1.5\textwidth]{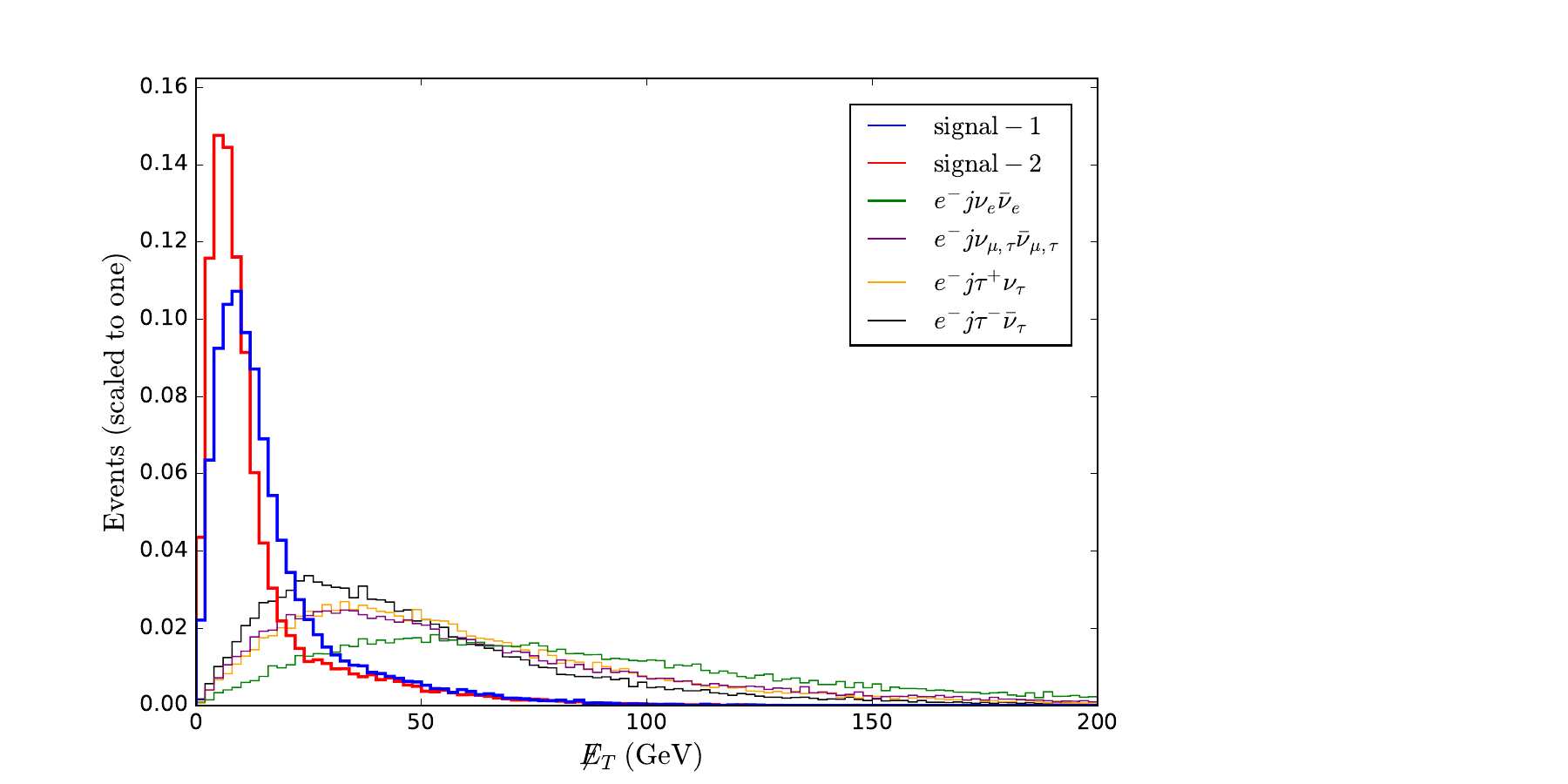}
		\end{minipage}
	}
	\subfigure[]{
		\begin{minipage}[b]{0.45\textwidth}
			\includegraphics[width=1.5\textwidth]{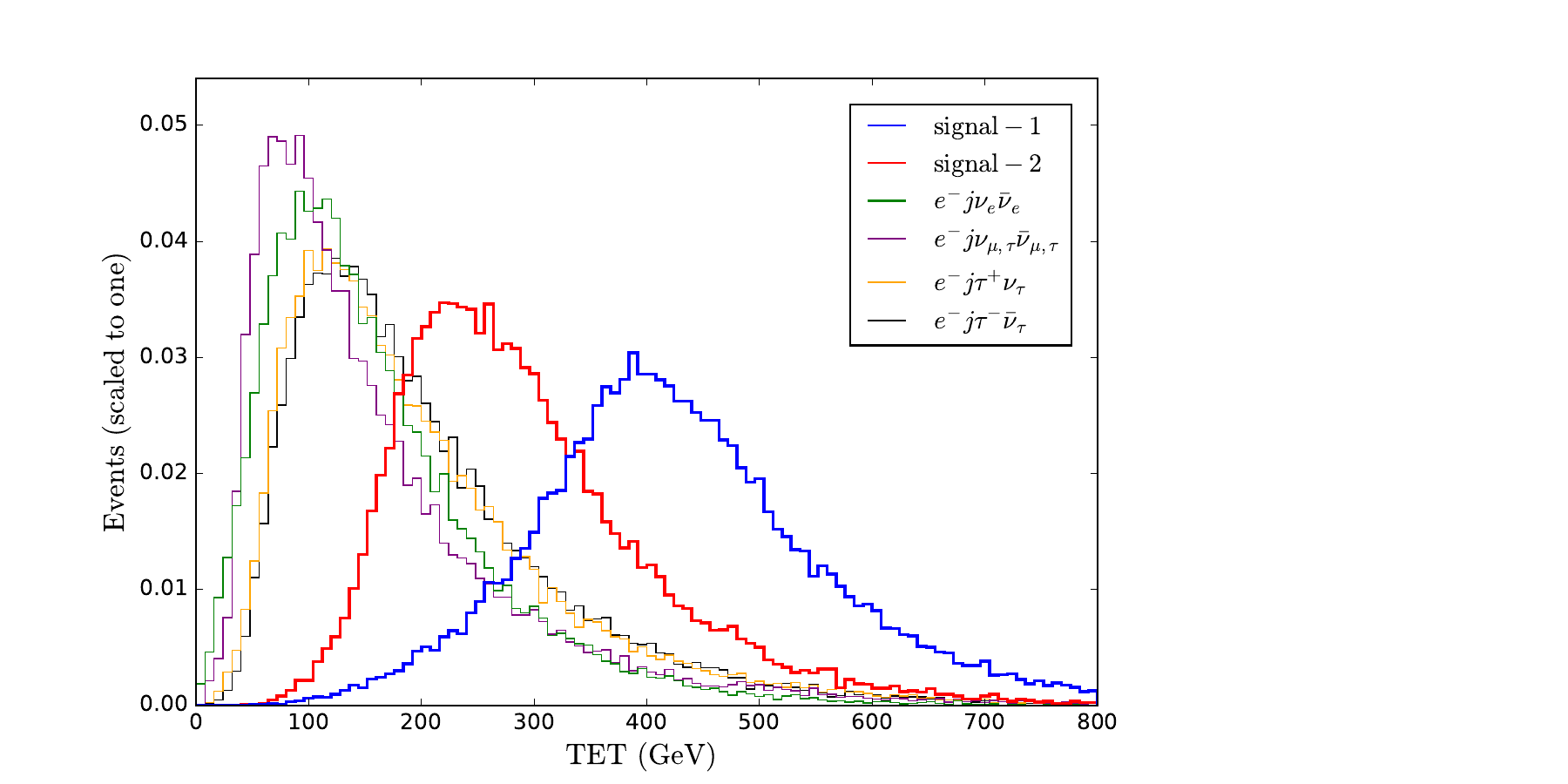}
		\end{minipage}
	}
	\centering
	\subfigure[]{	
		\begin{minipage}[b]{0.45\textwidth}
			\includegraphics[width=1.5\textwidth]{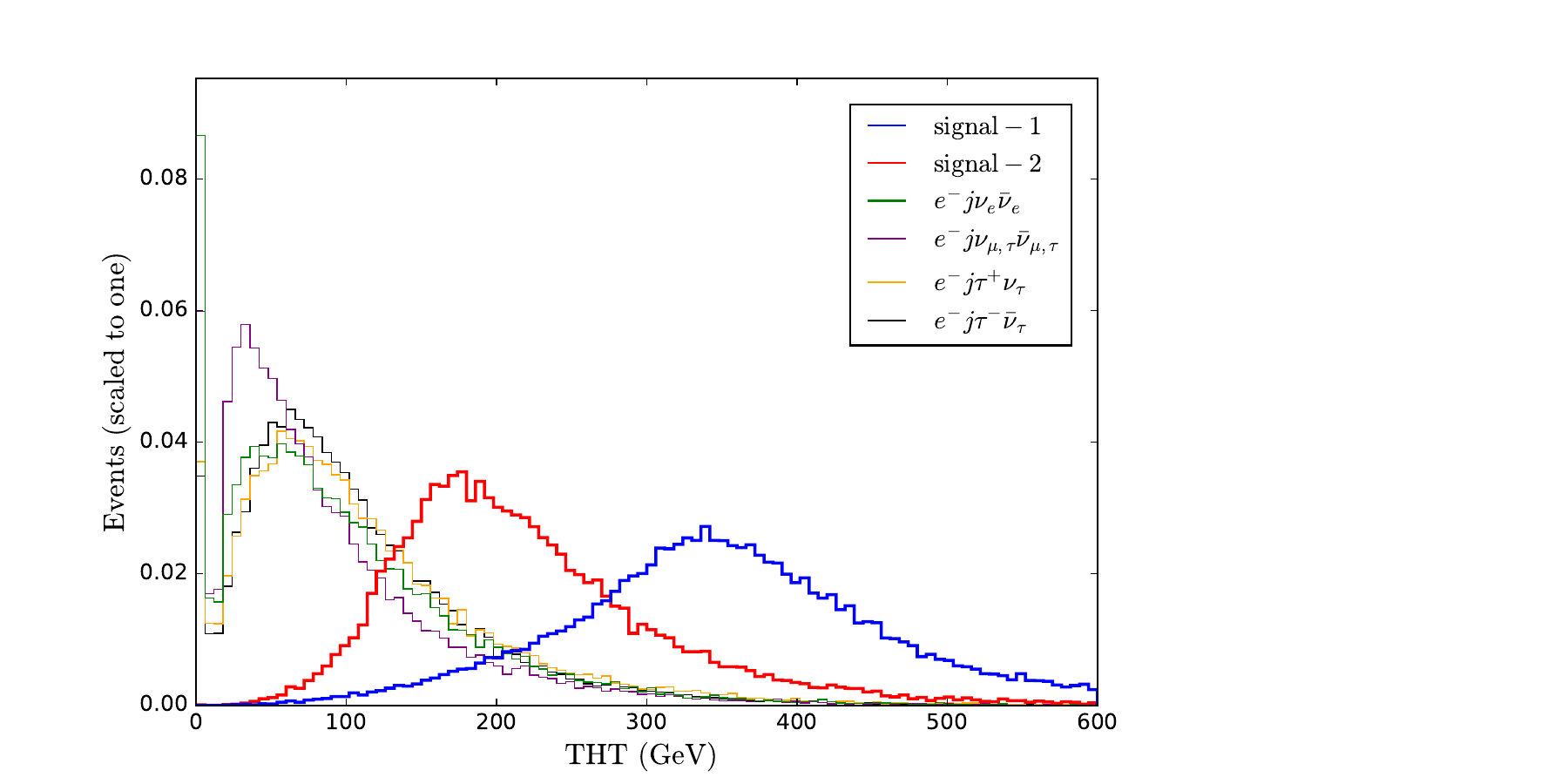}
		\end{minipage}
	}
	\centering
	\subfigure[]{	
		\begin{minipage}[b]{0.45\textwidth}
			\includegraphics[width=1.5\textwidth]{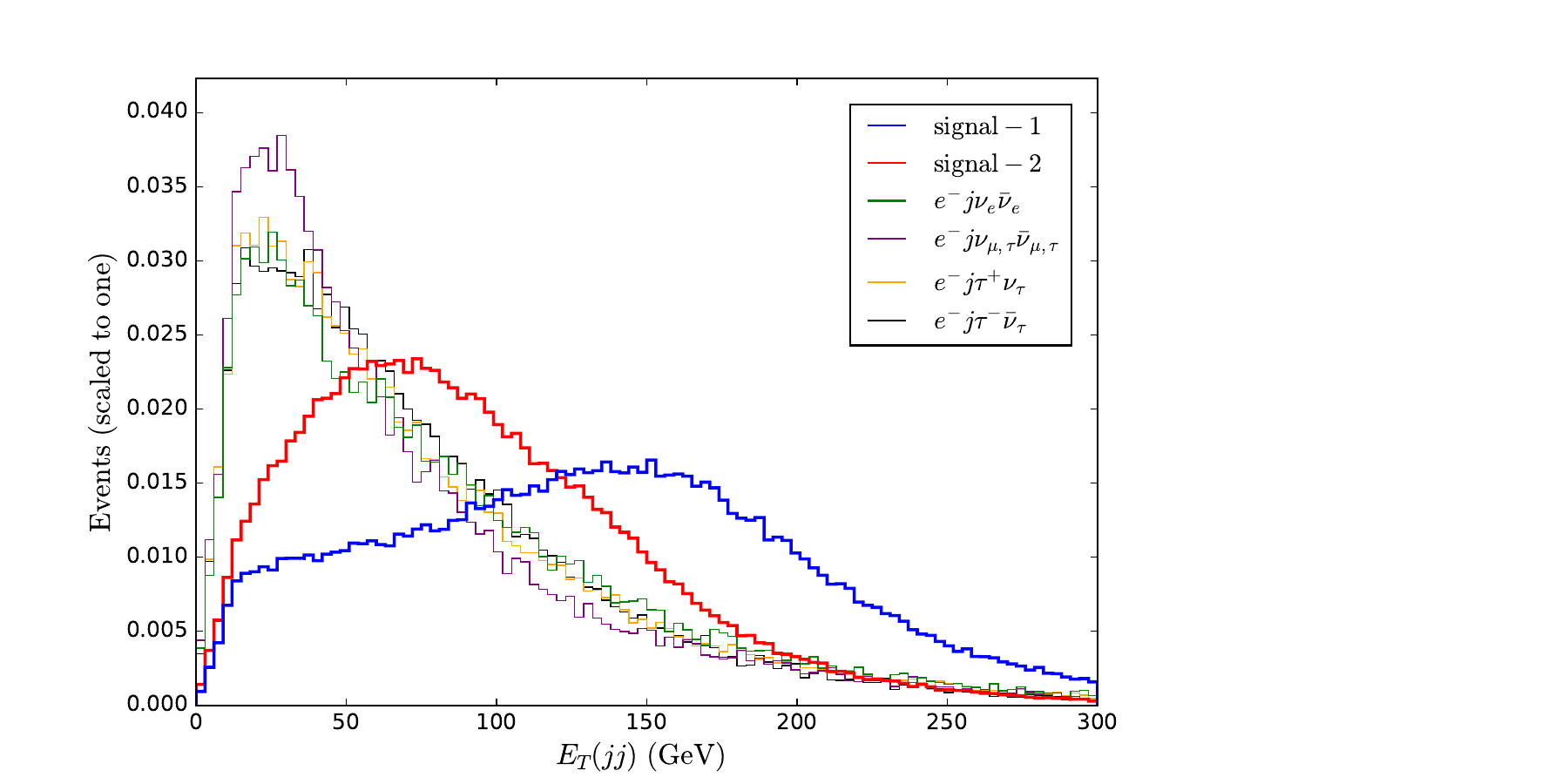}
		\end{minipage}
	}
	\centering
	\subfigure[]{	
		\begin{minipage}[b]{0.45\textwidth}
			\includegraphics[width=1.5\textwidth]{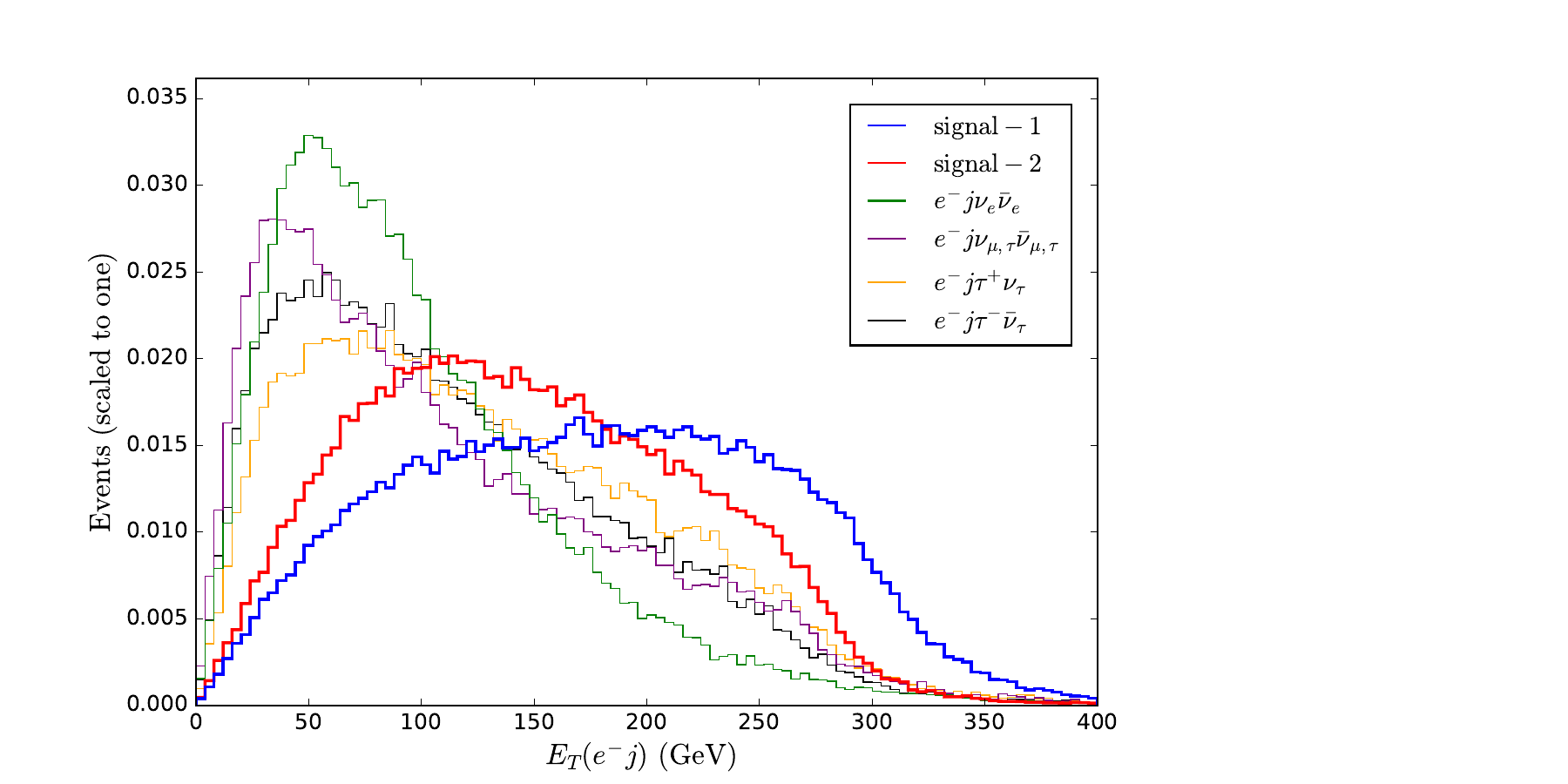}
		\end{minipage}
	}
	\caption{\footnotesize Normalized distributions of $\cancel{E}^{}_{T}$ (a), $\rm{TET}$ (b), $\rm{THT}$ (c), $E_{T}(jj)$ (d) and $E_{T}(e^{-}j)$ (e) for the signals and backgrounds at the LHeC with $E_{e^{-}}=140$~GeV and an integrated luminosity of 1 ab$^{-1}$.}
	\label{fig6}
\end{figure}
\begin{table*}[htbp]
	\caption{\footnotesize Effect of individual kinematical cuts on the signal-1 for ${M}_{h_{2}}=300$ GeV, ${M}_{Z^{}_{\mu\tau}}=0.1$ GeV, sin$\alpha=0.2$ and $g_{\mu\tau}=1\times10^{-3}$ and the SM backgrounds at the LHeC with $E_{e^{-}}=140~(60)$~GeV.  The statistical significance is computed for the integrated luminosity as 1 ab$^{-1}$.}
	\centering
	\begin{tabular}{c|c|c|c}
		\hline
		\hline
		\multicolumn{4}{c}{LHeC,\quad\quad $E_{e^{-}}=140~(60)$ GeV,\quad\quad $E_{p}=7$ TeV}\\
		\hline
		cuts                   & signal (S)     &  total background (B)& $S/\sqrt{S+B}$ \\
		\hline initial (no cut)     &~~~~~~~~417.0 (104.0)  &  $1.01\times10^{6}_{}$ ( $5.09\times10^{5}_{}$)  &  0.41 (0.15)  \\
		\hline basic cuts                     &~~~~~~~~392.1 (97.6)       &  $8.53\times10^{5}_{}$ ( $4.11\times10^{5}_{}$)  & 0.42 (0.16)  \\
		\hline $\cancel{E}^{}_{T} <20 $  ($\cancel{E}^{}_{T} <20 $) GeV & 238.6 (70.7)    &    $5.17\times10^{4}_{}$ ( $4.89\times10^{4}_{}$)           &1.05 (0.32) \\
		\hline $\rm{TET}>300$   ($\rm{TET} >260 $) GeV   & 235.9 (64.6)    &    $1.65\times10^{4}_{}$ ( $4.63\times10^{3}_{}$)           &1.83 (0.94) \\
		\hline $\rm{THT} >200 $ ($\rm{THT} >200 $) GeV  & 235.7 (63.8)    &    $9.91\times10^{3}$ ( $1.42\times10^{3}_{}$)           &2.30 (1.65) \\
		\hline $E_{T}(jj)> 100$ ($E_{T}(jj)> 90$) GeV   & 232.7 (60.4)   &   $7.83\times10^{3}_{}$ ( $1.08\times10^{3}_{}$)           &2.59 (1.79) \\
		\hline $E_{T}(e^{-}j)> 150$ ($E_{T}(e^{-}j)> 120$) GeV  & 232.3 (59.5)   &   $7.44\times10^{3}_{}$ ( $9.93\times10^{2}_{}$)           &2.65 (1.84) \\
		\hline
		\hline
	\end{tabular}
	\label{table2}
\end{table*}
\begin{table*}[htbp]
	\caption{\footnotesize Effect of individual kinematical cuts on the signal-2 for ${M}_{Z^{}_{\mu\tau}}=0.1$ GeV and $\chi=9\times10^{-5}~\rm{GeV^{-1}}$ and the SM backgrounds at the LHeC with $E_{e^{-}}=140~(60)$~GeV.  The statistical significance is computed for the integrated luminosity as 1 ab$^{-1}$.}
	\centering
	\begin{tabular}{c|c|c|c}
		\hline
		\hline
		\multicolumn{4}{c}{LHeC,\quad\quad $E_{e^{-}}=140~(60)$ GeV,\quad\quad $E_{p}=7$ TeV}\\
		\hline
		cuts                   & signal (S)     &  total background (B)& $S/\sqrt{S+B}$ \\
		\hline initial (no cut)     &~~~~~~~~1288.0 (544.0)  &  $1.01\times10^{6}_{}$ ( $5.09\times10^{5}_{}$)  &  1.28 (0.76)  \\
		\hline basic cuts                     &~~~~~~~~1205.9 (508.1)       &  $8.53\times10^{5}_{}$ ( $4.11\times10^{5}_{}$)  & 1.30 (0.79)  \\
		\hline $\cancel{E}^{}_{T} <20 $  ($\cancel{E}^{}_{T} <18 $) GeV & 980.9 (386.0)    &    $8.79\times10^{4}_{}$ ( $4.05\times10^{4}_{}$)           &3.29 (1.91) \\
		\hline $\rm{TET}>200 $  ($\rm{TET} >160 $) GeV   & 827.7 (361.9)    &    $2.26\times10^{4}_{}$ ( $1.27\times10^{4}_{}$)           &5.41 (3.16) \\
		\hline $\rm{THT} >140 $ ($\rm{THT} >120 $) GeV  & 808.1 (354.3)    &    $1.20\times10^{4}$ ( $6.09\times10^{3}_{}$)           &7.14 (4.41) \\
		\hline $E_{T}(jj)> 60$ ($E_{T}(jj)> 60$) GeV   & 803.5 (374.8)   &   $9.65\times10^{3}_{}$ ( $4.46\times10^{3}_{}$)           &7.86 (5.02) \\
		\hline $E_{T}(e^{-}j)> 100$ ($E_{T}(e^{-}j)> 80$) GeV  & 796.1 (343.7)   &   $8.96\times10^{3}_{}$ ( $4.19\times10^{3}_{}$)           &8.06 (5.10) \\
		\hline
		\hline
	\end{tabular}
	\label{table3}
\end{table*}

After the basic cuts, we further employ optimized kinematical cuts on separating the signals from the SM backgrounds. In our theoretical framework, although the SM backgrounds have a huge effect on the signals, there are many kinematical differences between them that can be exploited. In Fig.~\ref{fig6}, we show the normalized distributions of the total missing transverse energy $\cancel{E}^{}_{T}$, the visible transverse energy TET, the missing transverse hadronic energy THT, jet pair transverse energy $E_{T}(jj)$ and electron jet transverse energy $E_{T}(e^{-}j)$ for the signals and backgrounds at the LHeC with $E_{e^{-}}=140$~GeV and an integrated luminosity of 1 ab$^{-1}$. From these figures, we can see that the distributions of signals have good distinctions from the distributions of the relevant backgrounds (peaks locate in different locations). In principle, there are other variables which we can use to discriminate the signals from backgrounds. But, these variables are remarkably similar and can not work significantly better than above kinematic variables.
After all these kinematical cuts are applied, the event numbers of signal-1, signal-2 and corresponding backgrounds are summarized in Table~\ref{table2} and Table~\ref{table3} for the LHeC with $E_{e^{-}}=140~(60)$~GeV, respectively. The values of the statistical significance $SS$ are also shown in these tables, which is defined as $SS=S/\sqrt{S+B}$ with $S$ and $B$ being the number of signal and background events, respectively.
\begin{table*}[htbp]
	\caption{\footnotesize Effect of individual kinematical cuts on the signal-1 for ${M}_{h_{2}}=300$ GeV, ${M}_{Z^{}_{\mu\tau}}=0.1$ GeV, sin$\alpha=0.2$ and $g_{\mu\tau}=1\times10^{-3}$ and backgrounds at the FCC-eh. The statistical significance $SS$ is computed for an integrated luminosity of 1 ab$^{-1}$.}
	\centering
	\begin{tabular}{c|c|c|c}
		\hline
		\hline
		\multicolumn{4}{c}{FCC-eh,\quad\quad $E_{e^{-}}=60$ GeV,\quad\quad $E_{p}=50$ TeV}\\
		\hline
		cuts                   & signal (S)     &  total background (B)& $S/\sqrt{S+B}$ \\
		\hline initial (no cut)     &925.0  &  $1.81\times10^{6}_{}$   &  0.69\\
		\hline basic cuts                     &863.3     &  $1.30\times10^{6}_{}$   & 0.76 \\
		\hline $\cancel{E}^{}_{T} <20 $ GeV   & 303.2    &    $1.29\times10^{5}_{}$            &0.84  \\
		\hline $\rm{TET} >280 $ GeV    & 279.1     &    $1.35\times10^{4}_{}$           &2.38  \\
		\hline $\rm{THT} >200 $ GeV    & 277.7     &    $7.34\times10^{3}$            &3.18  \\
		\hline $E_{T}(jj)>100$ GeV  & 263.7    &   $5.84\times10^{3}_{}$            &3.38  \\
		\hline $E_{T}(e^{-}j)> 120$ GeV   & 261.9    &   $5.61\times10^{3}_{}$           &3.42 \\
		\hline
		\hline
	\end{tabular}
	\label{table4}
\end{table*}

\begin{table*}[htbp]
	\caption{\footnotesize Effect of individual kinematical cuts on the signal-2 for ${M}_{Z^{}_{\mu\tau}}=0.1$ GeV and $\chi=9\times10^{-5}~\rm{GeV^{-1}}$ and backgrounds at the FCC-eh. The statistical significance $SS$ is computed for an integrated luminosity of 1 ab$^{-1}$.}
	\centering
	\begin{tabular}{c|c|c|c}
		\hline
		\hline
		\multicolumn{4}{c}{FCC-eh,\quad\quad $E_{e^{-}}=60$ GeV,\quad\quad $E_{p}=50$ TeV}\\
		\hline
		cuts                   & signal (S)     &  total background (B)& $S/\sqrt{S+B}$ \\
		\hline initial (no cut)     &2412.0  &  $1.82\times10^{6}_{}$   &  1.79 \\
		\hline basic cuts                     &2246.7     &  $1.30\times10^{6}_{}$   & 1.96 \\
		\hline $\cancel{E}^{}_{T} <20 $ GeV   & 1086.9    &    $1.29\times10^{5}_{}$            &3.01  \\
		\hline $\rm{TET} >180 $ GeV    & 981.0     &    $3.76\times10^{4}_{}$           &5.00  \\
		\hline $\rm{THT} >120 $ GeV    & 972.0     &    $2.46\times10^{4}$            &6.08  \\
		\hline $E_{T}(jj)>60$ GeV  & 959.3    &   $1.92\times10^{4}_{}$            &6.76  \\
		\hline $E_{T}(e^{-}j)> 80$ GeV   & 952.9    &   $1.79\times10^{4}_{}$           &6.94  \\
		\hline
		\hline
	\end{tabular}
	\label{table5}
\end{table*}	

On the other hand, it is well known that the FCC-eh collides electrons to protons with $E_{e^{-}}=60$ GeV and $E_{p}=50$ TeV, which is a typical deep inelastic facility with
$\sqrt{s}\approx3.5$ TeV. Therefore, we need to modify the above veto criteria and kinematic cuts to adjust the progressive detector simulation, because the FCC-eh has a higher proton beam energy than LHeC. The modified values for the kinematic cuts and the event numbers of signal-1, signal-2 and backgrounds are presented in Table~\ref{table4} and Table~\ref{table5}, respectively.

\begin{figure}[!ht]
	\centering
	\subfigure[]{
		\begin{minipage}[b]{0.47\textwidth}
			\includegraphics[width=1\textwidth]{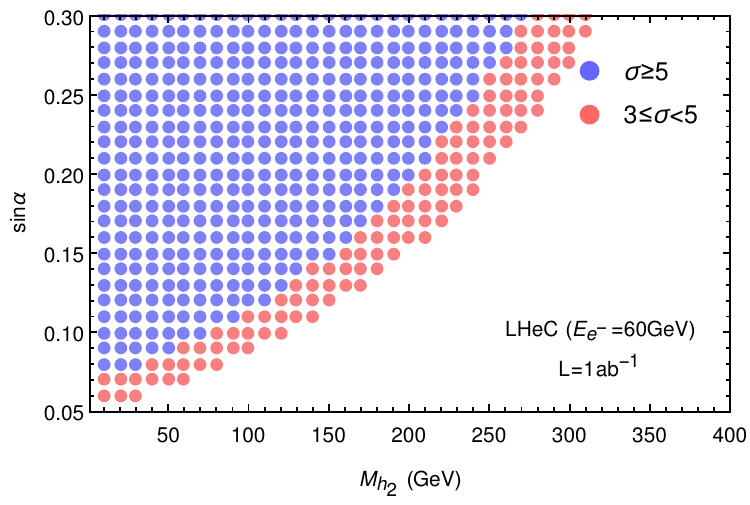}
		\end{minipage}
	}
	\subfigure[]{
		\begin{minipage}[b]{0.46\textwidth}
			\includegraphics[width=1\textwidth]{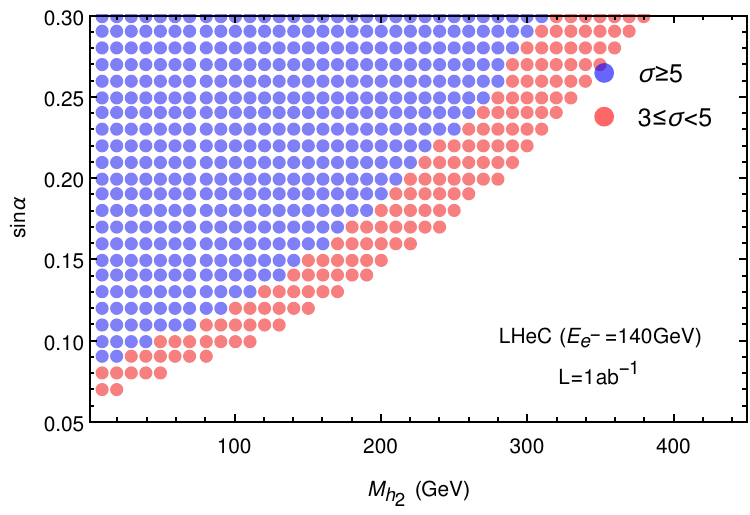}
		\end{minipage}
	}
	\subfigure[]{
		\begin{minipage}[b]{0.47\textwidth}
			\includegraphics[width=1\textwidth]{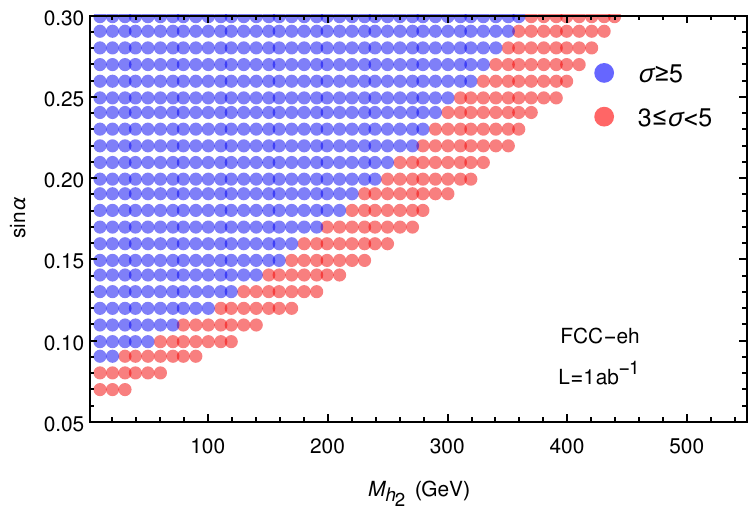}
		\end{minipage}
	}
	\caption{\footnotesize  3$\sigma$ and 5$\sigma$ detection potential regions for the signal-1 at the LHeC with $E_{e^{-}}=60$ GeV (a), LHeC with $E_{e^{-}}=140$ GeV (b) and FCC-eh (c) with an integrated luminosity of 1 ab$^{-1}$, respectively.}
	\label{fig7}
\end{figure}

  From these tables, we can see that, for sin$\alpha=0.2$, ${M}_{Z^{}_{\mu\tau}}=0.1$ GeV, $M_{h_{2}}=300$ GeV, $g_{\mu\tau}=1\times10^{-3}$ and the integrated luminosity being 1 ab$^{-1}$, the values of $SS$ for signal-1 can reach 2.65 (1.84) at the LHeC with $E_{e^{-}}=140~(60)$ GeV and 3.42 at the FCC-eh with  $E_{e^{-}}=60~\rm{GeV}$, $E_{p}=50 ~\rm{TeV}$. The values of $SS$ for signal-2 can reach 8.06 (5.10) at the LHeC with $E_{e^{-}}=140~(60)$ GeV and 6.94 at the FCC-eh with  $E_{e^{-}}=60~\rm{GeV}$, $E_{p}=50 ~\rm{TeV}$ when we take ${M}_{Z^{}_{\mu\tau}}=0.1$ GeV, $\chi=9\times10^{-5}~\rm{GeV^{-1}}$ and an integrated luminosity of 1 ab$^{-1}$.

 In Figs.~\ref{fig7} (a),~\ref{fig7} (b) and~\ref{fig7} (c), performing the scan over the parameter spaces of
${M}_{h^{}_{2}}$ and sin$\alpha$ for $M_{Z_{\mu\tau}}=0.1$ GeV, we show the experimental evidence region ($3\leq \rm{SS}< 5$) and experimental discovery region ($5\leq \rm{SS}$) of signal-1 at different e-p colliders with integrated luminosity being 1 ab$^{-1}$. For $\sin\alpha\leq0.3$, from Fig.~\ref{fig7} (a), we obtain the $h_{2}$ mass region of above $3\sigma$ confidence level as $10{~\rm{GeV}}\leq {M}_{h^{}_{2}}\leq320~\rm{GeV}$ and above $5\sigma$ confidence level as $10{~\rm{GeV}}\leq {M}_{h^{}_{2}}\leq270~\rm{GeV}$ at the LHeC with $E_{e^{-}}=60$ GeV. From Fig.~\ref{fig7} (b), we obtain the $h_{2}$ mass region of above $3\sigma$ confidence level as $10{~\rm{GeV}}\leq {M}_{h^{}_{2}}\leq400~\rm{GeV}$ and above $5\sigma$ confidence level as $10{~\rm{GeV}}\leq {M}_{h^{}_{2}}\leq310~\rm{GeV}$ at the LHeC with $E_{e^{-}}=140$ GeV. From Fig.~\ref{fig7} (c), we obtain the $h_{2}$ mass region of above $3\sigma$ confidence level as $10{~\rm{GeV}}\leq {M}_{h^{}_{2}}\leq480~\rm{GeV}$ and above $5\sigma$ confidence level as $10{~\rm{GeV}}\leq {M}_{h^{}_{2}}\leq360~\rm{GeV}$ at the FCC-eh. Based on these numerical results, we can say that the possible signatures of $h_{2}$ and $Z_{\mu\tau}$ from signal-1 is limited in the lower ${M}_{h^{}_{2}}$ range and could be detected at e-p colliders with an integrated luminosity of 1 ab$^{-1}$. On the other side, the FCC-eh could offer a better detection capabilities than LHeC under the same integrated luminosity.
\begin{figure}[!ht]
	\centering
	\subfigure{
		\begin{minipage}[b]{0.7\textwidth}
			\includegraphics[width=1\textwidth]{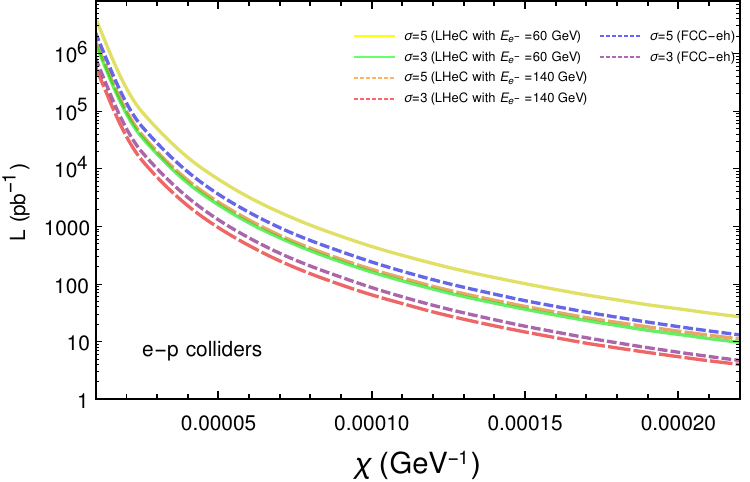}
		\end{minipage}
	}
	\caption{\footnotesize Integrated luminosity required for observing the signal-2 at the 3$\sigma$ and 5$\sigma$  statistical significance at different values of $\chi$ at e-p colliders.}
	\label{fig8}
\end{figure}
\begin{figure}[!ht]
	\centering
	\subfigure[]{
		\begin{minipage}[b]{0.47\textwidth}
			\includegraphics[width=1\textwidth]{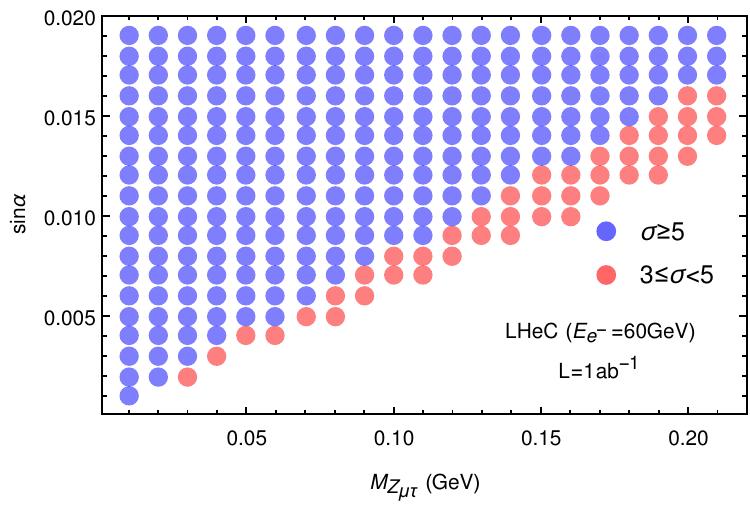}
		\end{minipage}
	}
	\subfigure[]{
		\begin{minipage}[b]{0.46\textwidth}
			\includegraphics[width=1\textwidth]{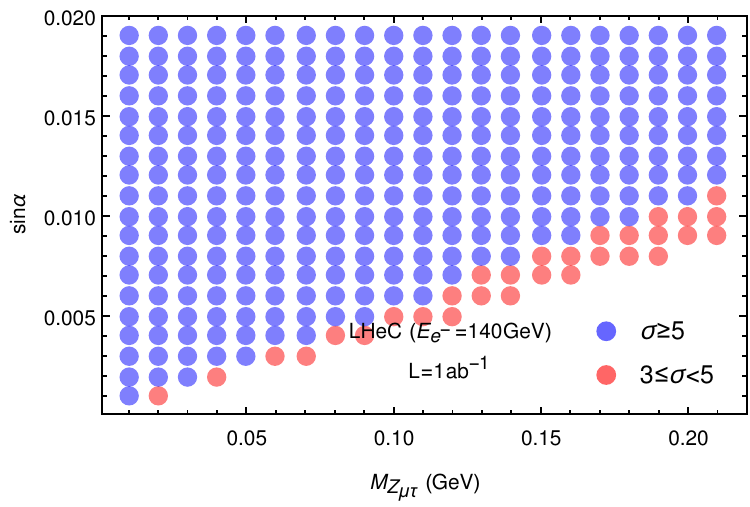}
		\end{minipage}
	}
	\subfigure[]{
		\begin{minipage}[b]{0.47\textwidth}
			\includegraphics[width=1\textwidth]{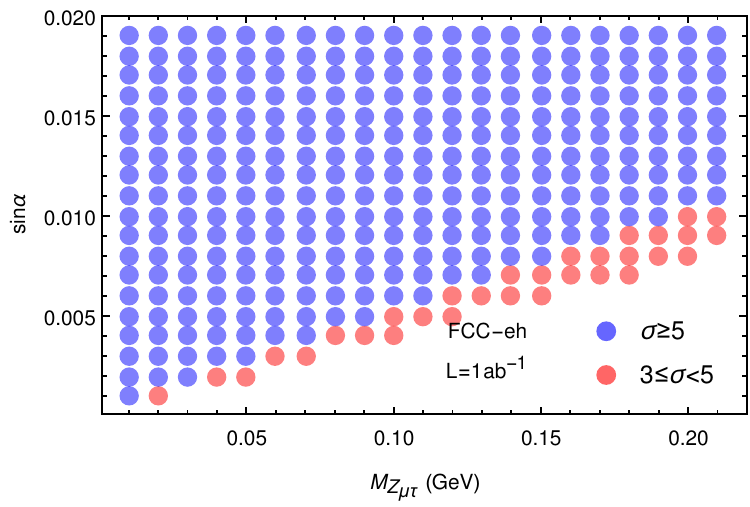}
		\end{minipage}
	}
	\caption{\footnotesize  3$\sigma$ and 5$\sigma$ detection  regions for the signal-2 at the LHeC with $E_{e^{-}}=60$ GeV (a), LHeC with $E_{e^{-}}=140$ GeV (b) and FCC-eh (c) with an integrated luminosity of 1 ab$^{-1}$, respectively.}
	\label{fig9}
\end{figure}

The required integrated luminosities for observing the new gauge boson $Z^{}_{\mu\tau}$ from signal-2 at the 3$\sigma$ and 5$\sigma$  confidence levels at different e-p colliders are plotted as functions of $\chi$ in Fig.~\ref{fig8}. We can see that one can obtain larger statistical significance for larger $\chi$ value within allowed parameter space. We can easily obtain $5\sigma$ statistical significance for taking $\chi\ge5\times10^{-5}$ within the designed luminosity region. Furthermore, the LHeC with $E_{e^{-}}=140$ GeV has the best sensitivity to the signal-2. However, it is likely that the properties of the SM-like Higgs boson will be narrowed before 1 ab$^{-1}$ of data are collected at an e-p collider. So we present Fig.~\ref{fig9} with no thought of $BR(h_{1}\to \rm{inv})$. In order to illustrate excluded regions of the free parameters ${M}_{Z^{}_{\mu\tau}}$ and sin$\alpha$ for reaching a given statistical significance, in Figs.~\ref{fig9} (a--b), we plot the $3\sigma$ and $5\sigma$ discovery of signal-2 for $g_{\mu\tau}=1\times10^{-3}$ in the sin$\alpha$-${M}_{Z^{}_{\mu\tau}}$ plane at different e-p colliders with an integrated luminosity 1 ab$^{-1}$. From these figures, we can see that one can easily obtain $5\sigma$ statistical significance for taking lower values of sin$\alpha$ in the parameter space of ${M}_{Z^{}_{\mu\tau}}$. Thus, from a phenomenological point of view, the signal-2 is more likely to result in a detection of the new gauge boson $Z^{}_{\mu\tau}$ at a lower integrated luminosity and more achievable experimental conditions.\\

 \textbf{ B. $h_{2}\to ZZ \to 4l$  Channel}\\

In this subsection we proceed to investigate the prospects of e-p colliders in searching for the scalar boson $h_{2}$ by focusing on its leading decay mode  $ZZ$
\begin{eqnarray}
e^{-}+p\to{\nu+j}+{h^{}_{ 2}(\to{ZZ})\to}~ 2l^{+}+2l^{-}+j+\cancel{E}^{}_{T}\;.
\end{eqnarray}
In above equation we have required the two $Z$ bosons to decay leptonically, then the signal consists of two lepton pairs, one jet and a large missing transverse energy $\cancel{E}^{}_{T}$.

For the signal $ 2l^{+}+2l^{-}+j+\cancel{E}^{}_{T}$, the main sources of the irreducible backgrounds come from the processes $e^{-}+p\to \nu jZZ$, $e^{-} jZZ$, $e^{-}jW^{+}W^{-}$, $e^{-} jZW^{+}$, $\nu jW^{+}W^{-}$ and $\nu jZW^{-}$. Similar as above we can  calculate the statistical significance easily for the luminosity of 1 ab$^{-1}$ at e-p colliders.  Our numerical results reveal that it is challenging to discover the $h_{2}$ signature through the $h_{2}\to ZZ \to 4l$ channel at the e-p colliders for $M_{h_{2}}$ = 200 (600) GeV, with $\sigma3$ ($\sigma5$) level statistical significance. We find that detecting this kind of signal at $3\sigma$ level requires the integrated luminosity be larger than
$10^{5}$ ab$^{-1}$ for the benchmark point defined by $M_{h_{2}}=200$~GeV, $\sin\alpha=0.2$ and $g_{\mu\tau}=1\times10^{-3}$, which extremely outreaches the designed luminosity. Thus, we do not show the relevant numerical results.

Certainly,  the scalar $h_{2}$ can also decay to the modes  $W^{+}W^{-}$,  $t\bar{t}$ and $h_{1}h_{1}$ as long as its mass is large enough. However, the $h_{2}\to WW \to \nu l2j$ or $h_{2}\to WW \to 2\nu 2l$ channel is much more difficult to reconstruct compared to the $h_{2}\to ZZ \to 4l$ channel due to the final state neutrino which escapes from the detector and makes it impossible to fully reconstruct the $h_{2}$ system. So we're not going to consider the channel here even though it has a larger branching ratio than that of $h_{2}\to ZZ $. We have to say that the detection of the new scalar boson $h_{2}$ via it decaying to pair of SM particles at e-p colliders is difficult to achieve at present. Of course, we also don't rule out the existence of some unique kinematic cuts that we do not take into account to optimize background suppression and improve signal observability.

\section*{\uppercase\expandafter{\romannumeral6}. Conclusions}
The ${U(1)}_{L^{}_{\mu}-L^{}_{\tau}}$  model, which can explain the muon $(g-2)$ anomaly, small neutrino masses and provide a candidate of DM, is phenomenologically rich and predictive. In this model, the additional scalar $h_{2}$ and  gauge boson $Z_{\mu\tau}$ are obtained after spontaneous breaking of $L^{}_{\mu}-L^{}_{\tau}$ symmetry. New scalar $h_{2}$ mixing with the SM-like Higgs boson is helpful to improve the precision of Higgs boson measurements. Furthermore, the gauge boson $Z^{}_{\mu\tau}$ possessing a mass around the MeV scale can explain the deficit of cosmic neutrino flux and resolve the problem of muon $(g-2)$ anomaly and relic abundance of DM simultaneously. So, studying these two new particles is of great significance for exploring this kind of new physics models.  In this paper, we have studied the possibility of searching for the new particles $h_{2}$ and $Z^{}_{\mu\tau}$ at e-p colliders. Since $Z^{}_{\mu\tau}$ can not couple  with the SM quarks and the first generation leptons, it is very difficult to be produced directly at colliders. So we consider its productions via decays of $h^{}_{1}$ and $h^{}_{2}$. Although the CC production of $h^{}_{1}$ and $h^{}_{2}$ have larger cross sections, their final states will generate mono-jet plus missing energy, which accidentally coincides with the DIS backgrounds. Therefore we focus on NC production channels $e^{-}p\to{e^{-}jh_{1}(\to{Z^{}_{\mu\tau}Z^{}_{\mu\tau}})\to}~e^{-}j+\cancel{E}^{}_{T}$ and $e^{-}p\to{e^{-}jh_{2}(\to{Z^{}_{\mu\tau}Z^{}_{\mu\tau}})\to}~e^{-}j+\cancel{E}^{}_{T}$, which provide good kinematic handles to distinguish the signals from the SM backgrounds. In addition to this, we also study the CC production of $h_{2}$ and further consider its non-invisible decays(e.g. $Z$ gauge boson) as complementary.

After giving the decay width expressions of several main decay channels of new scalar $h_{2}$, we calculate the production cross sections of the processes $e^{-}p\rightarrow e^{-}jh_{2}$ and $e^{-}p\rightarrow \nu_{e} jh_{2}$ with the beam polarization P($e^{-}$)= -0.8 in the context of the ${U(1)}_{L^{}_{\mu}-L^{}_{\tau}}$  model. The production cross sections of $Z^{}_{\mu\tau}$ are further calculated. Then, we investigate the observability of $h_{2}$ and $Z^{}_{\mu\tau}$ through the signal-1 from the process  $e^{-}p\to{e^{-}jh_{2}(\to{Z^{}_{\mu\tau}Z^{}_{\mu\tau}})\to}~e^{-}j+\cancel{E}^{}_{T}$ and the signal-2 from the process  $e^{-}p\to{e^{-}jh_{1}(\to{Z^{}_{\mu\tau}Z^{}_{\mu\tau}})\to}~e^{-}j+\cancel{E}^{}_{T}$  at e-p colliders with  1 ab$^{-1}$ integrated luminosity. After simulating the signals as well as the relevant backgrounds, and applying suitable kinematic cuts on the variables $\cancel{E}^{}_{T}$, $\rm{TET}$, $\rm{THT}$, $E_{T}(jj)$ and $E_{T}(e^{-}j)$, the values of the statistical significance $ SS $ for signal-1 can reach 2.65 (1.84)  at the LHeC with $E_{e^{-}}=140~(60)$ GeV and 3.42 at the FCC-eh with  $E_{e^{-}}=60~\rm{GeV}$, $E_{p}=50 ~\rm{TeV}$ when we take sin$\alpha=0.2$, $g_{\mu\tau}=1\times10^{-3}$ GeV, ${M}_{Z^{}_{\mu\tau}}=0.1~\rm{GeV}$ and ${M}_{h^{}_{2}}=300$ GeV. While for signal-2, its values can reach 8.06 (5.10)  at the LHeC with $E_{e^{-}}=140~(60)$ GeV and 6.94 at the FCC-eh when we take $\chi=9\times10^{-5}~\rm{GeV^{-1}}$ and ${M}_{Z^{}_{\mu\tau}}=0.1~\rm{GeV}$. Performing the scan over all parameter space, we find that the signals of $h_{2}$ and $Z_{\mu\tau}$ from signal-1 is limited in the lower ${M}_{h^{}_{2}}$ range and could be detected at e-p colliders with an integrated luminosity of 1 ab$^{-1}$. The signal of $Z_{\mu\tau}$ might be easily detected via signal-2 at e-p colliders, while the LHeC with $E_{e^{-}}=140$ GeV has the best sensitivity to signal-2. In the end, we analysis the signals of $h_{2}$ through the CC production channel via its decaying into a pair of gauge bosons. However, due to the interference of many backgrounds and the low number of events, searching for these kind of signals are harder to achieve at e-p colliders. Thus, we expect that the possible signals of the ${U(1)}_{L^{}_{\mu}-L^{}_{\tau}}$  model might be detected at future e-p colliders via $h_{2}\to Z^{}_{\mu\tau}Z^{}_{\mu\tau}$ and $h_{1}\to Z^{}_{\mu\tau}Z^{}_{\mu\tau}$ channels.

\section*{ACKNOWLEDGEMENT}
 This work was supported in part by the National Natural Science Foundation of China under Grant Nos.11875157, 11847303 and 11605081.

\end{document}